\input{psfig}
\documentclass[]{aa}
\usepackage{graphicx}
\usepackage{natbib}
\usepackage{color}
\begin{document}
\title{On the age of the magnetically active WW Psa and TX Psa members of the $\beta$ Pictoris association}
\author{S.\,Messina\inst{1},  R.\,Santallo\inst{2}, T.G.\,Tan\inst{3}, P.\,Elliott\inst{4,5}, G.\,A.\,Feiden\inst{6}, A.\,Buccino\inst{7}\thanks{Visiting Astronomer, Complejo Astron\'omico El Leoncito operated under agreement between the Consejo Nacional de Investigaciones Cient\'ificas y T\'ecnicas de la Rep\'ublica Argentina and the National Universities of La Plata, C\'ordoba and San Juan.}, P.\,Mauas\inst{7}, R.\,Petrucci\inst{8,9}, E.\,Jofr\'e\inst{8,9}}
\offprints{Sergio Messina}
\institute{INAF-Catania Astrophysical Observatory, via S.Sofia, 78 I-95123 Catania, Italy \\
\email{sergio.messina@oact.inaf.it}
\and   
Southern Stars Observatory, Pamatai, Tahiti, French Polynesia \\
\email{santallo@southernstars-observatory.org}
\and  
Perth Exoplanet Survey Telescope, Perth, Australia\\
\email{tgtan@bigpond.net.au}
\and
European Southern Observatory, Alonso de Cordova 3107,  Vitacura Casilla 19001, Santiago 19, Chile\\
\email{p210@exeter.ac.uk}
\and
School of Physics, University of Exeter, Stocker Road, Exeter, EX4 4QL
\and
Department of Physics and Astronomy, Uppsala University, Box 516, 751 20 Uppsala, Sweden\\
\email{gregory.a.feiden@gmail.com}
\and
Instituto de Astronom\'ia y F\'isica del Espacio (IAFE-CONICET), Buenos Aires, Argentina\\
\email{abuccino@iafe.uba.ar; pablo@iafe.uba.ar}
\and
Observatorio Astron\'omico de C\'ordoba, Laprida 854, X5000BGR, C\'ordoba, Argentina\\
\email{romina@oac.unc.edu.ar; jofre.emiliano@gmail.com}
\and
Consejo Nacional de Investigaciones Cient\'ificas y T\'ecnicas (CONICET),  Argentina\\
}

\date{}
\titlerunning{Rotation among $\beta$ Pic members}
\authorrunning{S.\,Messina et al.}
\abstract {There are a variety of  different techniques available to estimate the ages of pre-main-sequence stars. \rm Components of physical pairs, thanks to  their strict  coevality and  the mass difference, such as the binary system analysed in this paper, are best suited to test the effectiveness of these \rm different techniques. \rm}
{We consider the system WW Psa + TX Psa whose membership of the 25-Myr $\beta$ Pictoris association has been well established by earlier works. We investigate which   age dating technique \rm  provides the best agreement between  the age of the  system and that of the association. \rm} {We have photometrically monitored WW Psa and TX Psa and measured their rotation periods   as \rm P = 2.37\,d and P = 1.086\,d, respectively.   We have retrieved from the literature their Li equivalent widths and measured their effective temperatures and luminosities. \rm We investigate whether the ages of these stars derived using three independent techniques, that is based on rotation,  Li equivalent widths,  and the position in the HR diagram are consistent with  the age of the $\beta$ Pictoris association.} {We find that the rotation periods and the Li contents of both stars are consistent with \rm  the distribution of other bona fide members of the cluster. On the contrary, the isochronal fitting provides similar ages for both stars, but a factor of about four younger than the quoted age of the association, or about 30\% younger when the effects of magnetic fields are included.} {We explore the origin of the discrepant age inferred from isochronal fitting, including the possibilities that either the two components may be unresolved binaries or that the basic stellar parameters of both components are altered by enhanced magnetic activity. The latter is found to be the more reasonable cause, suggesting that age \bf estimates \rm  based on the Li content is more reliable than isochronal fitting for pre-main-sequence stars with pronounced magnetic activity.}
\maketitle
\rm

\section{Introduction}

\object{WW Psa} and \object{TX Psa} form an interesting  magnetically active, pre-main-sequence \rm pair in the $\beta$ Pictoris association. Thanks to their proximity to the Sun (d = 20.75\,pc) and to their angular separation ($\rho$ = 36$^{\prime\prime}$), both components are spatially resolved allowing measurement of their individual physical parameters.  
Considering that this system is a well established member of the $\beta$ Pictoris association, whose most recent age determination based on the Lithium Depletion Boundary (LDB) modeling is quoted
as 25$\pm$3\,Myr (\citealt{Messina16a}), we explore whether a similar age for both components of this system is also retrieved by other age determination techniques based, 
specifically, on the Lithium equivalent width (EW), the rotation period that we measured for both components, and the position in the HR diagram.
The circumstance that both components of the system are coeval but with different basic properties offers the advantage to put further constraints
to the age determination techniques.\\
WW Psa (\object{HIP\,112312},  \object{GJ 871.1\,A}, RA = 22:44:57.97, DEC = $-$33:15:01.7, J2000.0)  is a very active M4V dwarf 
(\citealt{Shkolnik09})  in the Piscis Austrinus  constellation at a  distance   d = 20.75$\pm$0.25\,pc as measured by GAIA
(GAIA collaboration \citeyear{Gaia16}). \rm
Strong Ca H\&K and
X-ray emission were detected, respectively, by \citet{Beers96} and by \citet{Thomas98} in the ROSAT All-Sky Survey 
(\object{1RXS\,224457.7-331506}).  \citet{Song02} and  \citet{Torres06} have not detected the Li line in their spectra of WW Psa. \rm \\
 TX Psa (\object{GJ 871.1\,B}, RA = 22:45:00.05, DEC = $-$33:15:25.8, J2000.0) is a very active M4.5V dwarf. Flare activity on this star was first 
detected by  \citealt{Kunkel72}, and moderate Ca H\&K emission subsequently reported by  \citet{Beers96}. 
TX Psa exhibits very large Li EW (450$\pm$20\,m\AA,  \citealt{Torres06}; 290$\pm$30\,m\AA,  \citealt{Song02}).\\ 
Measurements of the projected rotational velocity  (see Table\,1) indicate that both stars are fast rotators.  Indeed, WW Psa has
a photometric rotation period P = 2.3546\,d (\citealt{Pojmanski97};  \citealt{Messina10}),  as inferred from the analysis of the ASAS (All Sky Automated Survey) photometric time series and TX Psa has a rotation period P = 1.086\,d as measured in the present study.
 \rm
 Fast rotation, enhanced magnetic activity, high Li abundance (in the case of TX Psa)
all suggest that these are very young stars. \\
 \citet{Song02},  on the basis of common proper motions, first proposed that WW Psa and TX Psa
may be components of a binary system with angular separation $\rho$ = 36$^{\prime\prime}$  ($\sim$747\,AU) and also members
of the young $\beta$ Pictoris association (\citealt{Song03}). Membership was    subsequently suggested by  \citet{Zuckerman04}, 
 \citet{Torres06}, \rm and finally confirmed
 by  \citet{Shkolnik12}. 
  Using the GAIA parallax and average radial velocities (see Table 1) we computed updated Galactic space and velocity components (see Table 1) that
  are consistent with those expected for $\beta$ Pictoris members (see, e.g.,  \citealt{Mamajek14}),
  consolidating the fact that these stars are bona fide members of the  $\beta$ Pictoris association.\rm \\
 A summary of the physical properties of WW Psa and TX Psa retrieved from the literature and derived in this work is given in Table\,1.\\
  One aspect that makes this system very interesting is the fact that its components straddle the Li depletion boundary of the $\beta$ Pictoris association (e.g.,  \citealt{Binks14};  \citealt{Malo14b};  \citealt{Messina16a}) with the primary component WW Psa almost completely depleted  and the secondary component TX Psa  on the Li un-depleted side of the boundary (Li EW = 450\,m\AA). \rm
Another interesting aspect concerns the age. In fact, the   isochronal \rm  age inferred by  \citet{Song02}, as well as that estimated in this work,  are significantly younger than the age inferred from the ensemble members of the \object{$\beta$ Pic association}, e.g.,  21$\pm$9 Myr (\citealt{Mentuch08}),  24$\pm$4\,Myr (\citealt{Binks16}), 22$\pm$3\,Myr (\citealt{Mamajek14}), 26$\pm$3\,Myr (\citealt{Malo14b}), or 25$\pm$3\,Myr (\citealt{Messina16a}). \\
To use  age-dating tecnhiques based on rotation and Li EW, which is also affected by rotation, and to compare the calculated ages with the above mentioned estimates,
we have carried out a photometric monitoring campaign to measure the rotation period of TX Psa, and to confirm the already known rotation period of WW Psa.

\begin{table*}
\caption{Physical properties of WW Psa and TX Psa}
\begin{tabular}{lccc}
\hline
Parameter & WW Psa & TX Psa & note\\
\hline
V (mag) & 12.10$\pm$0.02 & 13.35$\pm$0.02 &  1\\
V$-$K$_s$  (mag) & 5.16$\pm$0.03 & 5.51$\pm$0.03 & 2\\
Sp. Type & M4V & M4.5V & 3\\
Dist. (pc) &   20.75$\pm$0.25 &    20.75$\pm$0.25 & 4\\
$<$RV$>$ (km\,s$^{-1}$) & 2.42$\pm$0.84 & 2.54$\pm$0.60 & 1\\ 
X, Y, Z    (pc) & 9.48,    2.10,   -18.34 &  9.48,   2.10,  -18.34 & 1 \\
U, V, W (km\,s$^{-1}$) & -10.46$\pm$0.40,     -16.04$\pm$0.21,    -9.98$\pm$0.75 &  -10.41$\pm$0.31,    -16.03$\pm$0.20,    -10.09$\pm$0.54 & 1 \\
$v\sin{i}$ (km\,s$^{-1}$) & 12		&	22 & 5\\
				     & 12.1		&	16.8  & 6\\
				     & 14$\pm$1.73		&24.3$\pm$4.93 & 7\\
				     & 13.9$\pm$0.5		&22.7$\pm$0.3 & 8\\
T$_{\rm eff}$ (K) & 3200$\pm$100 & 3050$\pm$100 & 1 \\
M (M$_\odot$) &  0.19$\pm$0.04 & 0.11$\pm$0.03 & 1\\
L (L$_\odot$) &   0.046$\pm$0.006 &   0.021$\pm$0.003 & 1\\
R (R$_\odot$) & 0.70$\pm$0.13 & 0.52$\pm$0.10 & 1 \\
i ($^{\circ}$) &    53$^{+22}_{-12}$ &   60$^{+20}_{-18}$ & 1\\
Li EW (m\AA) & no Li & 450 & 9,10\\
P (d)  & 2.37$\pm$0.01 & 1.086$\pm$0.003 & 1\\
age (Myr) (Li) (Feiden models\rm) &\multicolumn{2}{c}{29$\pm$2} & 1\\
age$^a$ (Myr) (Feiden isochr.) &\multicolumn{2}{c}{$<$ 20} & 1\\
age$^b$ (Myr) (Feiden isochr.) &\multicolumn{2}{c}{27$\pm$7} & 1\\
\hline

\multicolumn{4}{l}{1:  this work; 2: 2MASS;  3: \citet{Shkolnik09} 4: GAIA collaboration (\citeyear{Gaia16}); 5: \citet{Christian02};}\\
\multicolumn{4}{l}{ $6$:  \citet{Torres06}; $7$: \citet{Jayawardhana06}; $8$:  \citet{Bailey12}; $9$: \citet{Song02}; $10$: \citet{Torres06};  }\\
\multicolumn{4}{l}{$^a$: if single stars; $^b$: if unresolved binaries.} \\
\end{tabular}
\end{table*}

\section{Observations}
We carried out photometric monitoring of WW Psa and TX Psa in 2014 and 2015 at three different observatories.

\subsection{Siding Spring Observatory (SSO)}

We observed the system with a 32-cm f/7.4 Ritchey-Chretien telescope at the Siding Spring Observatory
(31$^{\circ}$\,16$^{\prime}$\,24$^{\prime\prime}$S,  149$^{\circ}$\,03$^{\prime}$\, 52$^{\prime\prime}$E, 1165\,m a.s.l., Australia).
The telescope is equipped with a ST8-XME CCD camera (9$\mu$m pixels size, and a plate scale of 0.8$^{\prime\prime}$/pixel),
has a corrected 13.6$^{\prime}$$\times$20.4$^{\prime}$ field of view and mounts BVR filters. Our photometric monitoring was carried out 
from September 19 to November 2, 2014 for a total of 21 nights. On each night we collected on average three consecutive frames in the V filter for a total
of 57 frames using an integration time of 180 sec per frame. We used  tasks within IRAF\footnote{IRAF is distributed by the National Optical Astronomy Observatory, which 
	is operated by the association of the Universities for Research in Astronomy, inc. (AURA) under cooperative agreement with the National 
	Science Foundation.}  for bias
correction and flat fielding, and the technique of aperture photometry to extract magnitude time series for
the targets and for other stars detected in the frames,    that were selected  as candidate comparison stars.\\ \rm
  We identified two stars  that were found to be non variable and were used to build an ensemble comparison (C1 and C2 in  Table 2). \rm
 We measured a standard deviation $\sigma_{\rm C1-C2}$ = 0.006\,mag in their differential light curve   over the 21 nights. \rm
The magnitudes of WW Psa and TX Psa were computed differentially with respect to the ensemble comparison.

After averaging the three consecutive differential magnitudes obtained on each night (collected within a time interval of about
15 minutes), we obtained  21 average V-band differential magnitudes for each component of the binary system for the subsequent analysis. The average standard 
deviation associated   with \rm  the nightly averaged magnitudes is $\sigma_V$ = 0.004\,mag, which we consider as our photometric precision.

\begin{table*}
\caption{Properties of the selected comparison stars\label{comp}}
\begin{tabular}{llccc}
   \hline
   & name & RA (J2000) & DEC (J2000)   & V mag\\
   \hline
C1 & \object{TYC\,7501\,1270\,1} & 22:45:17.60 & $-$33:10:43.65  &    13.79$^{1}$\\

C2 & \object{2MASS\,J22454487-3313241} & 22:45:44.87 	& $-$33:13:24.19  &   12.54$^{2}$ \\

C3 & \object{TYC\,7501\,1081\,1}  & 22:45:40.68 &  $-$33:18:30.85 & 12.50$^3$ \\
\hline
\multicolumn{5}{l}{$^1$: from \citet{Hog00}; $^{2}$: determined in the present work;}\\
\multicolumn{5}{l}{$^3$:  APASS (Munari et al. 2014)}\\
\end{tabular}
\end{table*}

\subsection{CASLEO}
We had two additional  nights of observation, 18 and 20 November, 2014, at the   Complejo
Astron\'omico El Leoncito (CASLEO) Observatory (31$^{\circ}$\,47$^{\prime}$\,57$^{\prime\prime}$S,  
69$^{\circ}$\,18$^{\prime}$\, 12$^{\prime\prime}$W, 2552\,m a.s.l., San Juan, Argentina). We used the Horacio Ghielmetti Telescope (THG)
which is a 40-cm f/8 remotely-operated MEADE-RCX 400 Ritchey-Chretien telescope,
equipped with an Apogee Alta U16M camera and Johnson UBVRI and Clear filters. The CCD has $4096 \times
4096$ pixels, 9$\mu$m pixel size, a plate scale of 0.57$^{\prime\prime}$/pixel, and
a 49$^{\prime}\times$49$^{\prime}$ field of view (\citealt{Petrucci13}).
We used the B, V, R, and I filters and collected a total of 37 frames per filter. Data reduction  and magnitude extraction \rm were carried out as
outlined   in Sect.\,2.1.  
Only V filter data were used together with V filter data from SSO for the subsequent analysis.
 The standard 
deviation of the averaged nightly magnitudes is $\sigma_V$ = 0.004\,mag, which we consider as our photometric precision.

\begin{figure*}
\begin{minipage}{18cm}
\includegraphics[width=90mm,height=140mm,angle=90,trim= 0 0 0 0]{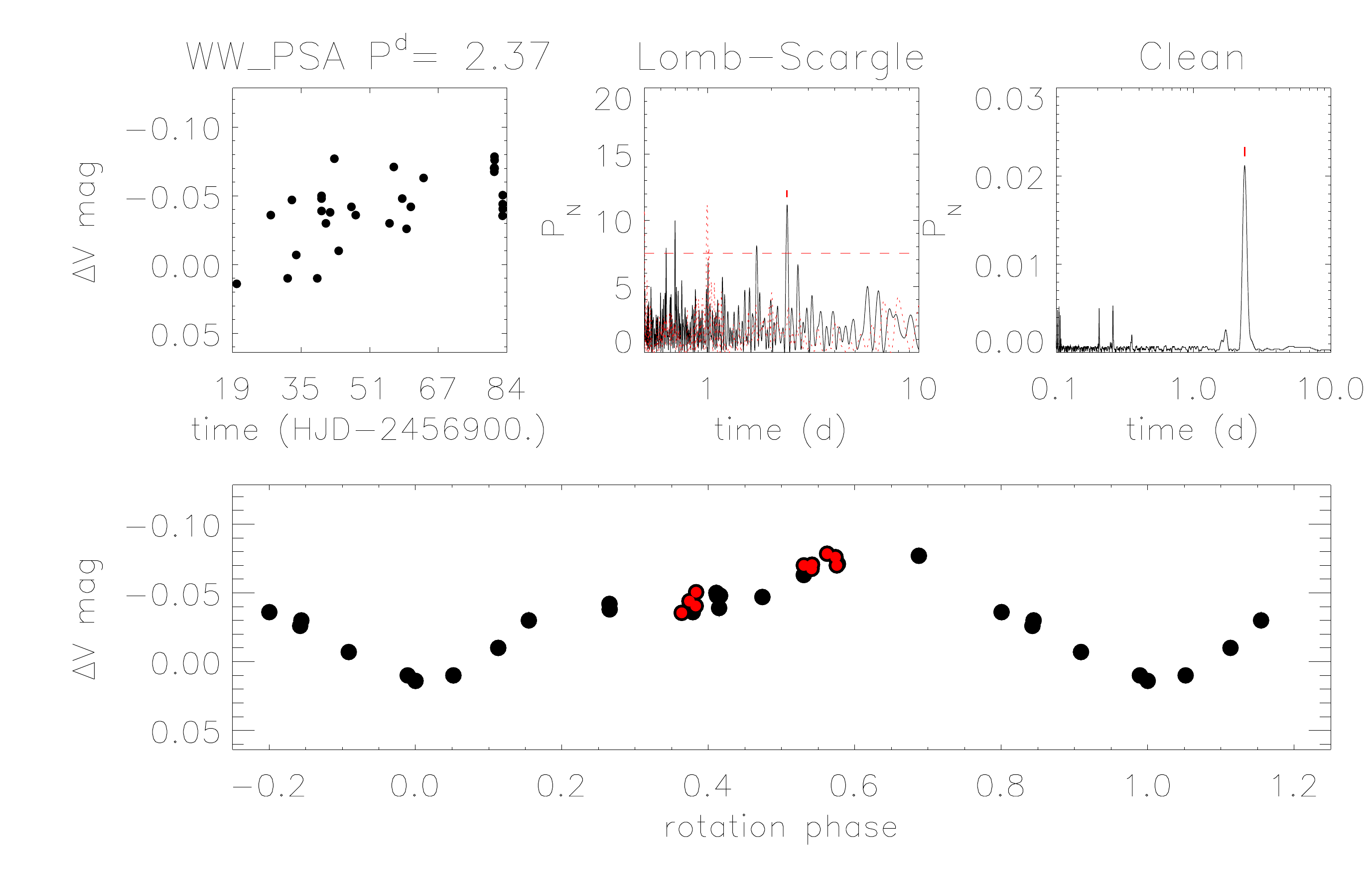} 
\end{minipage}
\caption{\label{wwpsa_sso}\it Top-left panel\rm: V-band magnitude timeseries of WW Psa versus Heliocentric Julian Day collected at Siding Spring Observatory and CASLEO. \it Top-middle panel: \rm Lomb-Scargle periodogram (solid line).   The power peak  corresponding to the rotation period P = 2.37\,d is marked with  a small red line above it. \rm  The dotted red line indicates the window spectral function, whereas the horizontal dashed line represents the power level corresponding to a FAP = 0.01. 
\it Top-right panel\rm: Clean periodogram. \it Bottom panel: \rm Light curve phased with the rotation period. Red bullets are data from   CASLEO. \rm The uncertainty associated  with \rm  each point is smaller than the symbol size. 
}
\end{figure*}  

\begin{figure*}
\begin{minipage}{18cm}
\includegraphics[width=90mm,height=140mm,angle=90,trim= 0 0 0 0]{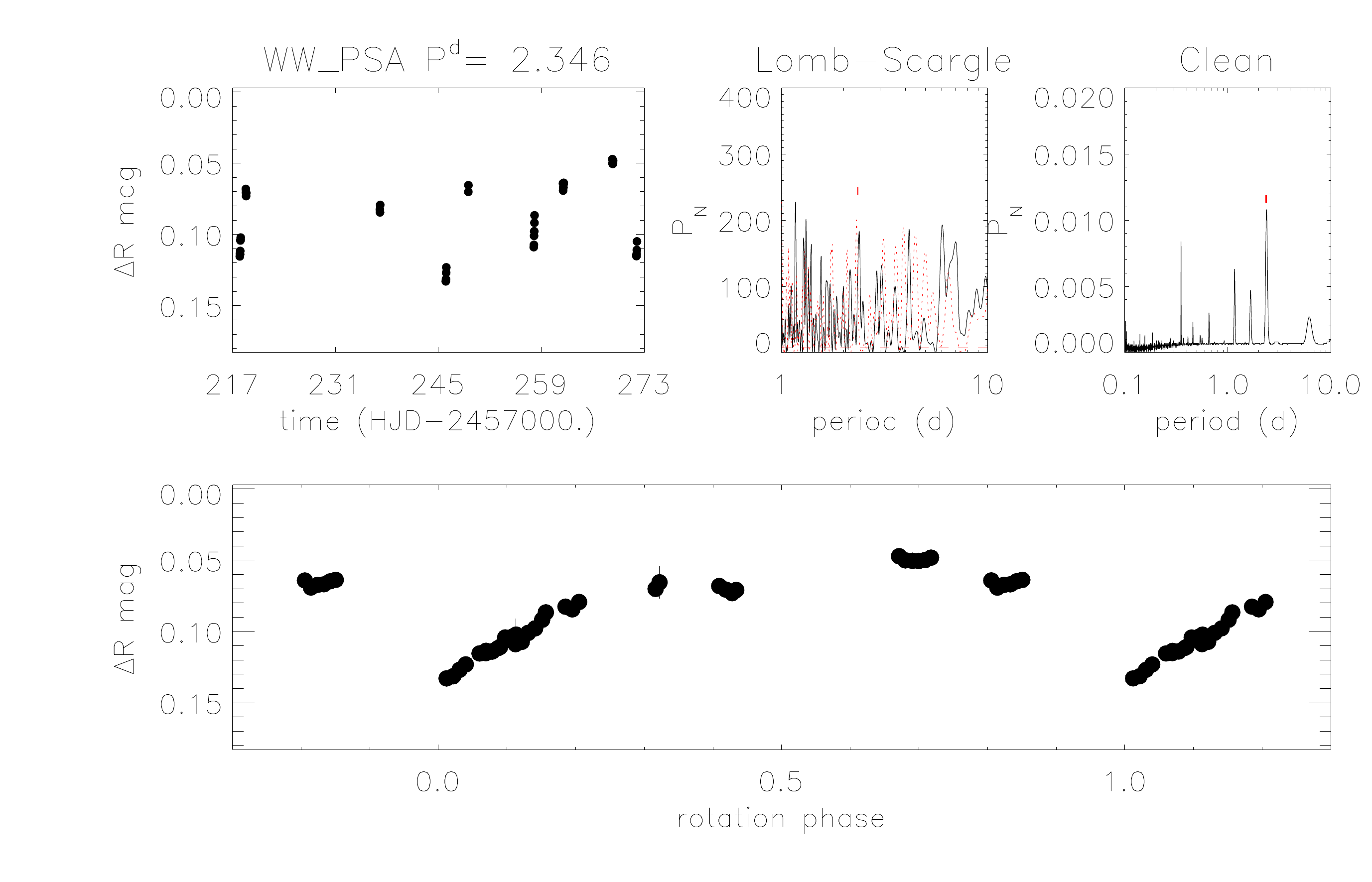}
\end{minipage}
\caption{\label{wwpsa_pest}Same as Fig.\,\ref{wwpsa_sso}, but with data collected at PEST Observatory.
}
\end{figure*}

\subsection{PEST Observatory}
The most recent data were collected at the PEST (Perth Exoplanet Survey Telescope) Observatory  (31$^{\circ}$\,58$^{\prime}$\, S, 115$^{\circ}$\,47$^{\prime}$\, E, 24\,m a.s.l., Perth, Australia).
We used  the 30-cm f/10 Meade LX200 SCT telescope with focal reducer yielding f/5 equipped with  a  SBIG  ST-8XME CCD camera and  a filter wheel loaded with BV(RI)$_{\rm c}$ and Clear filters.  
Focusing  was  computer  controlled with an Optec TCF-Si focuser.  The image scale obtained is 1.2$^{\prime\prime}$/pixel and the field of view is  31$^{\prime}$$\times$21$^{\prime}$. 
 Observations were collected  with the R$_{\rm c}$ filter using 120s integration time.  Observations were carried out from July 14 to September 7, 2015 for a total of 9 nights during which we achieved  a total of
 593 frames. On each night WW Psa and TX Psa were observed for up to nine consecutive hours.   Differently than SSO and CASLEO, frame reduction (dark subtraction and flat-fielding) and aperture photometry were carried out using an automatic pipeline based on C-Munipack (ver. 2.0.16) and optimized for PEST data. \rm
After averaging  consecutive magnitudes (collected within a time interval of about
30 minutes) we obtained  48 average R-band magnitudes for the subsequent analysis. The average standard 
deviation is $\sigma_{\rm R c}$ = 0.0025 mag for both targets, which we consider as our photometric precision.
Differential magnitudes of the targets were obtained using an ensemble comparison consisting of   C1, C2, and C3 (see Table 2). \rm

\section{Rotation period search}
\subsection{Photometric data}
\begin{figure*}
\begin{minipage}{18cm}
\includegraphics[width=90mm,height=140mm,angle=90,trim= 0 0 0 0]{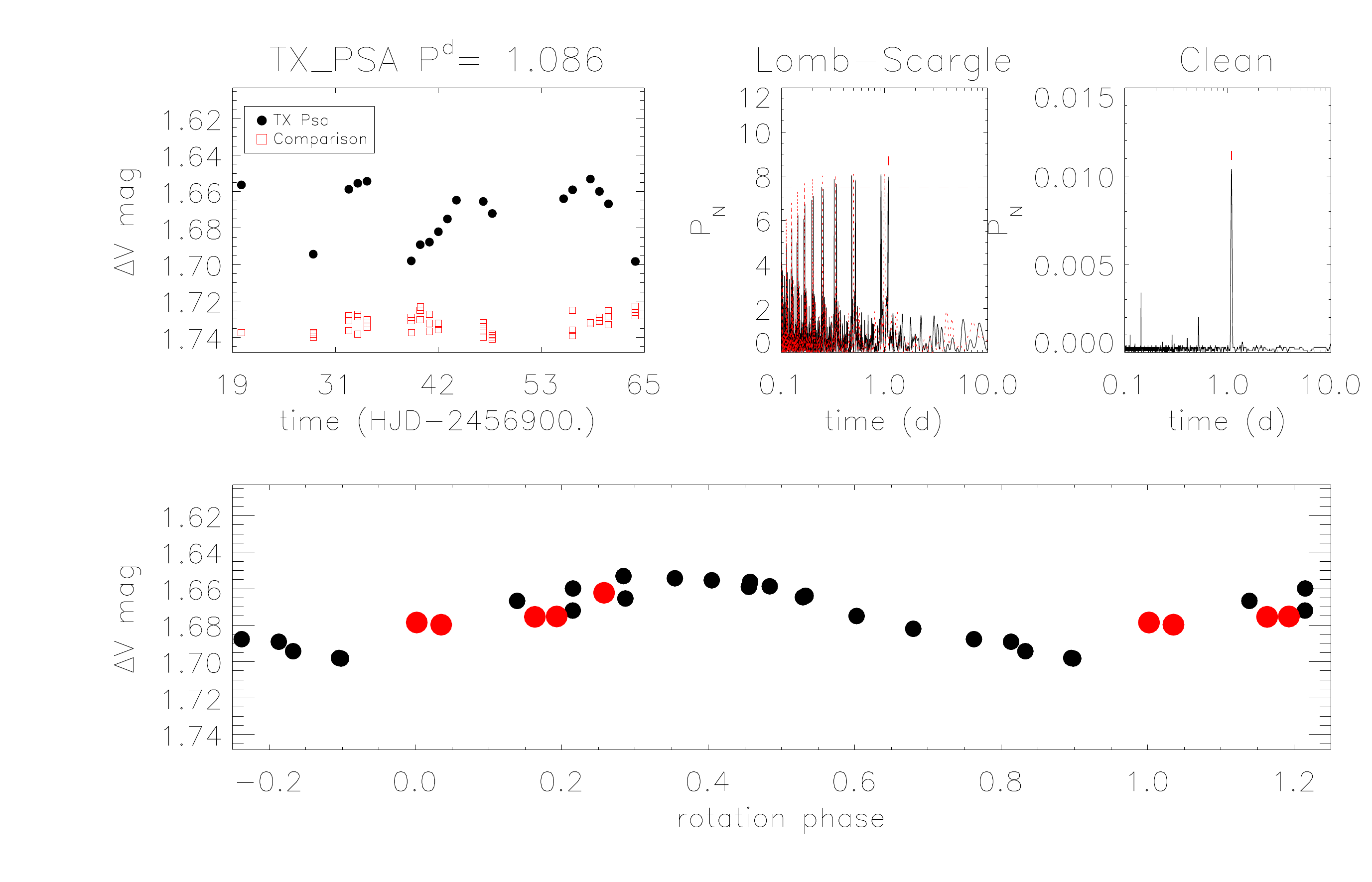}
\end{minipage}
\caption{\label{txpsa_sso}\it Top-left panel\rm: V-band  photometric \rm timeseries of TX Psa (bullets) versus Heliocentric Julian Day.    Red open squares represent the C1$-$C2 time series. \rm \it Top-middle panel: \rm Lomb-Scargle periodogram (solid line).  The power peak  corresponding to the rotation period P = 1.086\,d is marked with a small red line above it. \rm The dotted red line indicates the window spectral function, whereas the horizontal dashed line represents the power level corresponding to a FAP = 0.01. 
\it Top-right panel\rm: Clean periodogram. \it Bottom panel: \rm Light curve phased with the rotation period. Red bullets are data from   CASLEO. \rm}
\end{figure*}

\begin{figure*}
\begin{minipage}{18cm}
\includegraphics[width=90mm,height=140mm,angle=90,trim= 0 0 0 0]{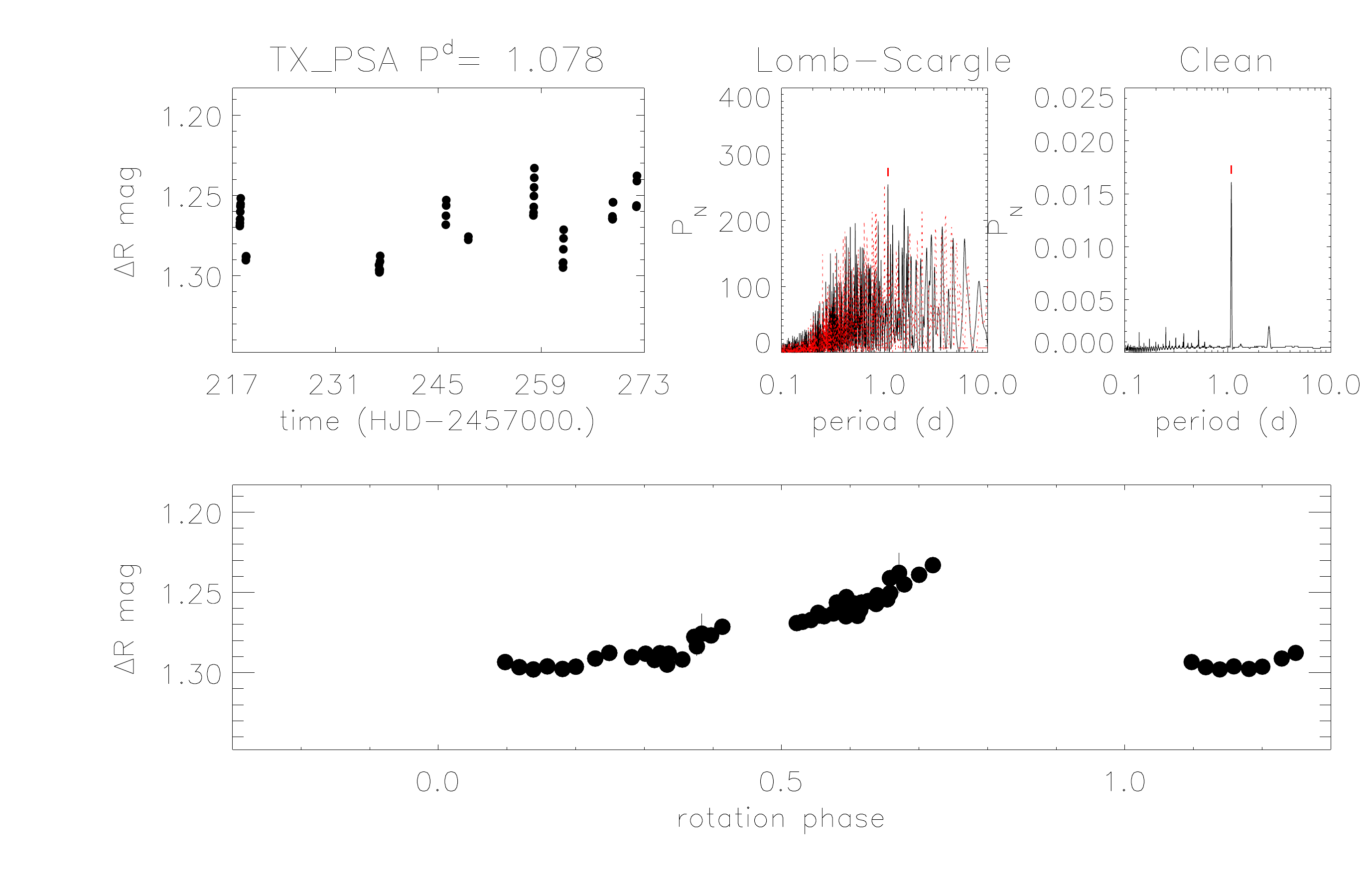} 
\end{minipage}
\caption{\label{txpsa_pest}Same as Fig.\,\ref{txpsa_sso}, but with data collected at PEST Observatory.}
\end{figure*}

We used the Lomb-Scargle (\citealt{Scargle82}) and Clean (\citealt{Roberts87}) periodogram analyses to search for significant periodicities in the WW Psa and TX Psa   photometric \rm  timeseries
related to their stellar rotation periods.   Use of two independent approaches  allows us to be more confident in the interpretation of the results of the periodogram analysis. \rm \\
The results of our analysis for WW Psa   and TX Psa \rm are plotted in Fig.\,\ref{wwpsa_sso} and Fig.\,\ref{txpsa_sso}, respectively, \rm for the data collected at SSO and CASLEO in 2014 and in  Fig.\,\ref{wwpsa_pest} and Fig.\,\ref{txpsa_pest}, respectively, \rm for the data collected with PEST in 2015. In the top-left panels we plot the  photometric \rm timeseries
versus the Heliocentric Julian Day (HJD), in the top-middle panel we plot the normalized Lomb-Scargle periodogram, where the red dotted line represents  the spectral window function   relative to \rm  the data sampling, and  the horizontal dashed line indicates the power level corresponding to a False Alarm Probability FAP = 1\%, which is the probability that a power peak of that height simply arises from Gaussian noise in the data. The FAP was estimated using a Monte-Carlo method, i.e., by generating 1000 artificial light curves obtained from the real one, keeping the date but   permuting \rm  the magnitude values. In the top-right panel, we plot the Clean periodogram where the power peak arising from the light rotational modulation dominates, whereas all secondary peaks, arising from the aliasing, are effectively removed. For WW Psa, our analysis finds that the stellar rotation period is P = 2.37$\pm$0.01\,d in the 2014 data time series.   Following the prescription of \citet{Lamm04}, the uncertainty in the period can be written as
\begin{displaymath}\Delta P = \frac{\delta \nu P^2}{2}, \end{displaymath} 	
where $\delta\nu$ is the finite frequency resolution of the power spectrum and is equal to the full width at half maximum of the main peak of the window function w($\nu$)   computed together with the LS periodogram and P is the the rotation period found by the LS periodogram. \rm If the time sampling is not too non-uniform, which is the case related to our observations, then $\delta\nu \simeq 1/T$, where T is the total time span of the observations.  \rm
In the bottom panel, we plot the light curve phased with the rotation period. 
The light curve   of WW Psa \rm  has a peak-to-peak amplitude  $\Delta$V = 0.09 mag and shows two light minima arising from active regions separated in longitude. The uncertainty associated with each point is smaller than the symbol size.\\
In Fig.\,\ref{wwpsa_pest}, we show the results for WW Psa obtained from the data collected with PEST in 2015. In this case, our analysis finds that the stellar rotation period is P = 2.346$\pm$0.005\,d, which is similar to the one found  from the data collected \rm in 2014. The peak-to-peak light curve amplitude in the R band is $\Delta$R = 0.08\,mag. The small difference between the two rotation periods may arise either from the effect of active regions growth and decay or from presence of surface differential rotation.

Our analysis of the TX Psa   photometric time series \rm finds that the stellar rotation period is P = 1.086$\pm$0.003\,d in  the data collected in \rm 2014.  In Fig.\,\ref{txpsa_sso},  we note that the power peak corresponding to the rotation period is separated by the 1-d peak of the spectral window function, and  is preserved
after cleaning by the CLEAN algorithm. 
The light curve of TX Psa in 2014 had a peak-to-peak amplitude  of $\Delta$V = 0.04\,mag.

In Fig.\,\ref{txpsa_pest} we show the results for TX Psa obtained from the data collected with PEST in 2015. In this case, our analysis finds that the stellar rotation period is  P = 1.078$\pm$0.006\,d, which is in agreement with the one found in 2014. 
  The phased light curve from PEST shows that data is not available at some parts of the phase.  So $\Delta$R might actually be larger than 0.07\,mag. \rm
The comparison between this peak-to-peak 
amplitude and that measured in 2014 ($\Delta$V = 0.03\,mag\footnote{We   note \rm that the amplitude of spot induced photometric variability in the V-band is larger than that in the R-band.}) denotes a significant increase in the activity level.

\subsection{Spectroscopic data}
\begin{figure*}
\begin{minipage}{18cm}
\includegraphics[width=90mm,height=140mm,angle=90,trim= 0 0 0 0]{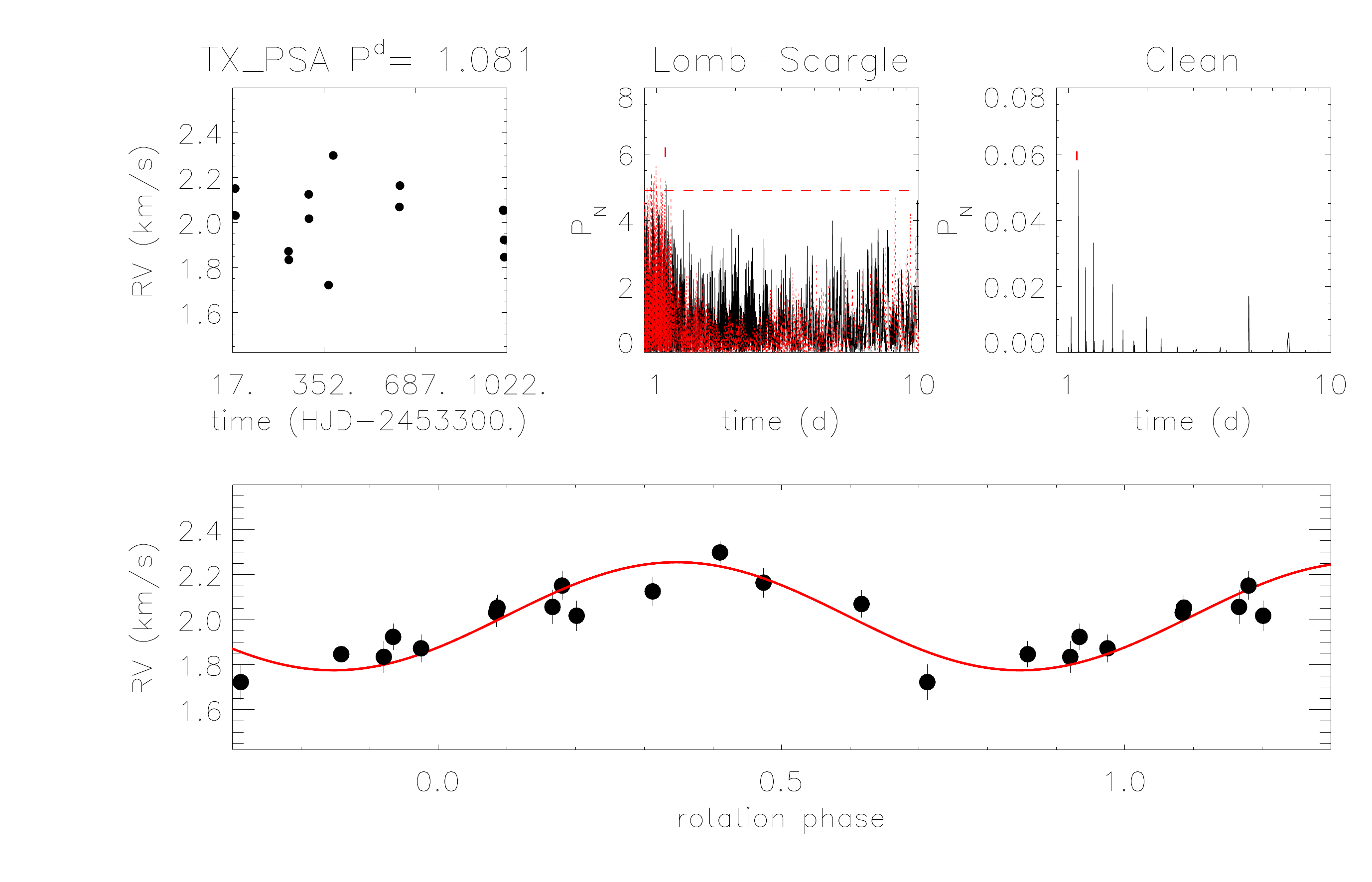} 
\end{minipage}
\caption{\label{txpsa_rv}\it Top-left panel\rm: Radial Velocity of TX Psa versus Heliocentric Julian Day. \it Top-middle panel: \rm Lomb-Scargle periodogram (solid line) with indication of the power peak corresponding to the rotation period P = 1.080\,d. The dotted red line indicates the window spectral function, whereas the horizontal dashed line represents the power level corresponding to a FAP = 0.02. 
\it Top-right panel\rm: Clean periodogram. \it Bottom panel: \rm  RV curve phased with the rotation period. The red solid line represents the sinusoidal fit with the rotation period. }
\end{figure*}

We retrieved a series of high-precision radial velocity 
measurements of both components from \citet{Bailey12}. 
The average values are $<$RV$>$ = 3.09$\pm$0.13\,km\,s$^{-1}$  and $<$RV$>$ = 2.03$\pm$0.17\,km\,s$^{-1}$ for WW Psa and TX Psa respectively. The measured dispersions have the same order of magnitude as those arising from magnetic activity
jitters at infrared wavelengths   (\citealt{Bailey12}). \rm We performed Lomb-Scargle and Clean periodogram analyses of these series. 
Whereas, in the case of WW Psa, our analysis did not find any evidence of significant periodicity at  confidence
levels larger than 80\%; in the case of TX Psa, we detected a power peak at P = 1.081$\pm$0.005 at 99\% confidence level.
When RV values are phased with this period (see Fig.\,\ref{txpsa_rv}), we find a peak-to-peak amplitude of 0.48\,km\,s$^{-1}$ against an average
precision of 0.06\,km\,s$^{-1}$. The similarity between this period and the rotation period (actually, in the case of   Lomb-Scargle \rm the
RV and photometric periods are equal), suggests that the radial velocity variation is induced by the stellar activity.
  
\section{The targets' physical properties}

\begin{figure*}
\begin{minipage}{18cm}
\includegraphics[width=60mm,height=80mm,angle=90,trim= 0 0 0 0]{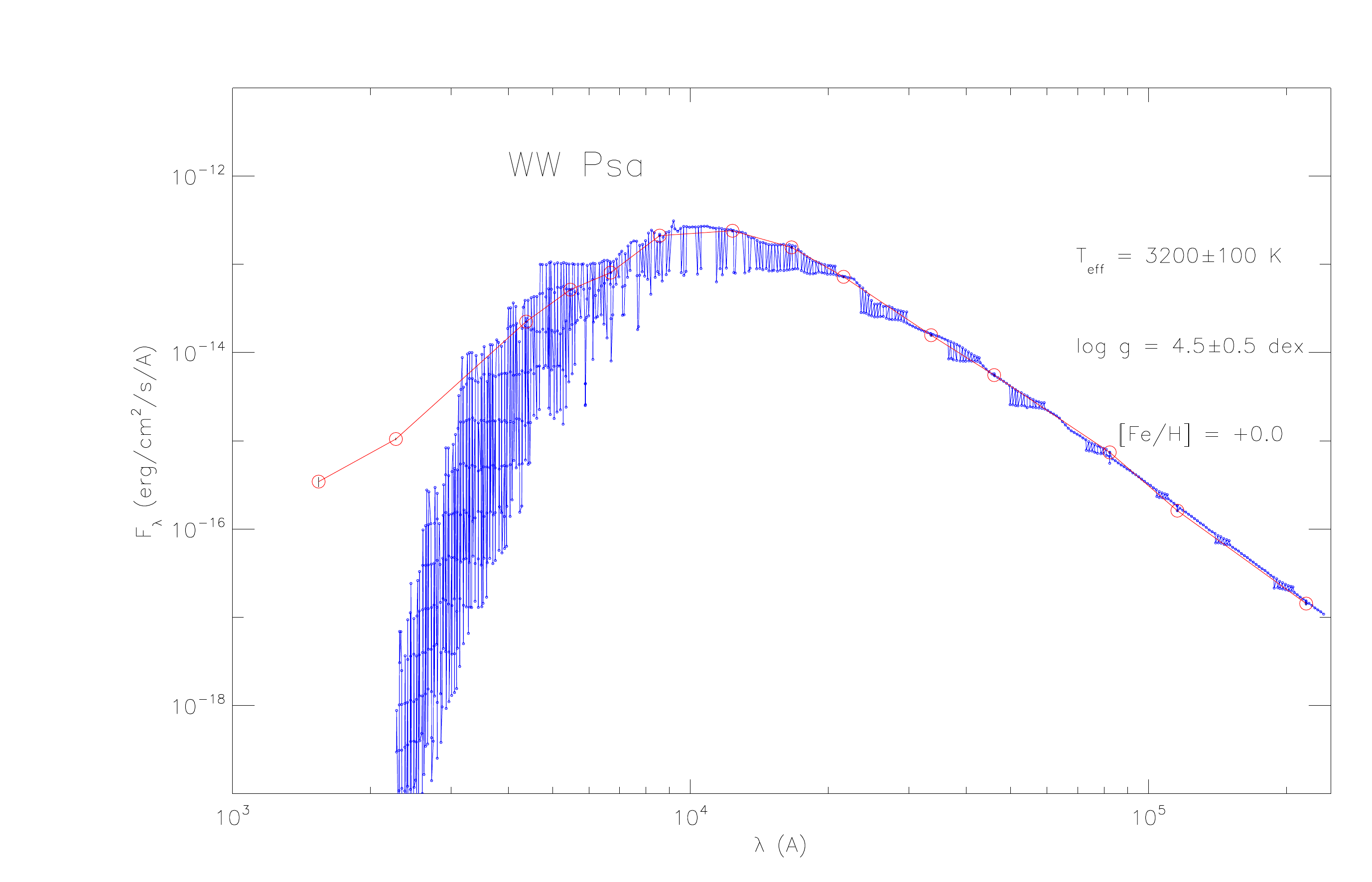} 
\includegraphics[width=60mm,height=80mm,angle=90,trim= 0 0 0 0]{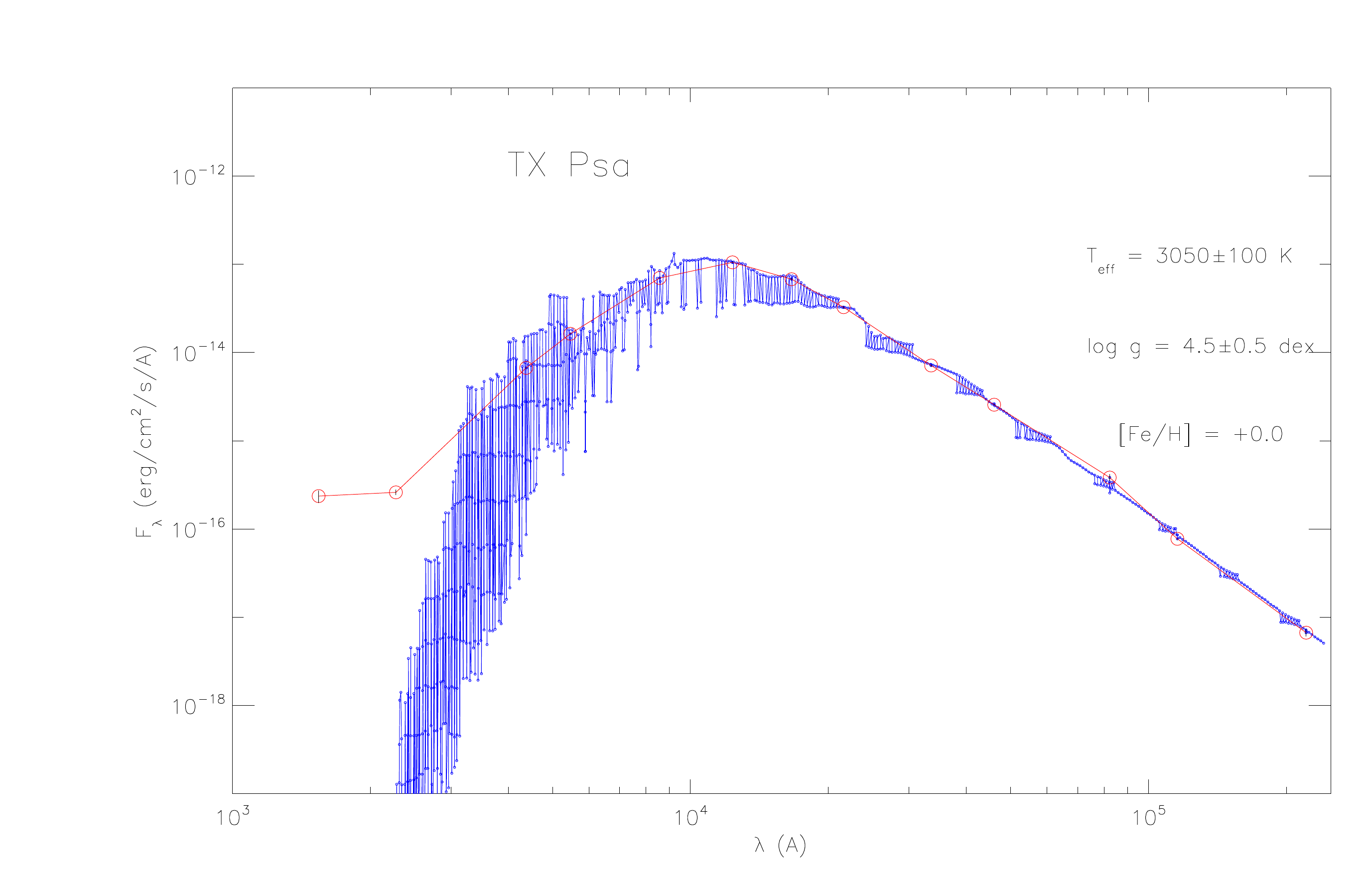} 
\end{minipage}
\caption{\label{sed}Spectral Energy Distributions  of WW Psa and TX Psa. Red bullets are the observed fluxes, whereas the blue line represents   the BT-NextGen best fit models as found by VOSA analysis.\rm }
\vspace{0cm}
\end{figure*}

We used literature optical    and IR \rm photometry to build the observed spectral energy distribution (SED).
In the case of WW Psa, we retrieved Near-UV and Far-UV magnitudes from GALEX (\citealt{Bianchi11}), UBVRI magnitudes from \citet{Koen10}, 
JHK magnitudes from the 2MASS  point source catalogue \rm (\citealt{Cutri03}), and W1--W4 magnitudes from   ALLWISE \rm  (\citealt{Cutri13}).
In the case of TX Psa, we retrieved Near-UV and Far-UV magnitudes from GALEX (\citealt{Bianchi11}), BVRI magnitudes from \citet{Casagrande08}   and \citet{Koen10},  \rm JHK magnitudes from the 2MASS   point source catalogue \rm  (\citealt{Cutri03}), and W1--W4 magnitudes from   ALLWISE \rm (\citealt{Cutri13}). \\
The SEDs were fitted  using the Virtual Observatory SED Analyser (VOSA; \citealt{Bayo08}) \rm with a grid of theoretical spectra from the   BT-NextGen \rm Model (\citealt{Allard12})  with effective temperatures ranging from  2700\,K \rm to 4000\,K, surface gravities from log g = 3\,dex to log = 5\,dex and fixing the metallicity to the average value [Fe/H] = +0.0\,dex observed among other members of the same association (see, e.g., \citealt{Mentuch08}) using solar abundances revised by \citet{Asplund09}. \rm The best fit by means of a $\chi^2$ minimization \rm is obtained with a model of  T$_{\rm eff}$ = 3200$\pm$100\,K and log g = 4.5$\pm$0.5 dex for WW Psa, \rm whereas for TX Psa with a model of   T$_{\rm eff}$ = 3050$\pm$100\,K and log g = 4.5$\pm$0.5 dex \rm (see Fig.\ref{sed}).  
For both targets we find no evidence of IR excess, whereas a significant Far- and Near-UV flux excess is observed.
Owing to their excess, fluxes in the U, Near-UV, and Far-UV bands were excluded from the fitting procedure.\\ \rm

\section{Age of the system}
In recent years a number of studies have provided estimates for the age of the $\beta$ Pictoris association
(see, \citealt{MacDonald10}, \citealt{Malo14b}, \citealt{Mamajek14}, \citealt{Binks16}, \citealt{Messina16a}).
The ages estimated from these studies range from 21 to 26\,Myr, with the most recent estimate of 25$\pm$3\,Myr by \citet{Messina16a} based on the LBD modeling after decorrelating the Li EW from the effects of stellar rotation. Only the estimate of 40\,Myr  by \citet{MacDonald10}
significantly deviates from the average.\\
In this study we intend to explore if the ages of both WW Psa and TX Psa are consistent with the $\beta$ Pictoris association average of 25$\pm$3 Myr based on three independent age-dating techniques; namely Li EW, rotation rates and position in the HR diagram. \\ \rm 

\subsection{Age from Lithium abundance}
\begin{figure*}
\begin{minipage}{20cm}
\includegraphics[width=50mm,height=80mm,angle=90,trim= 0 0 0 0]{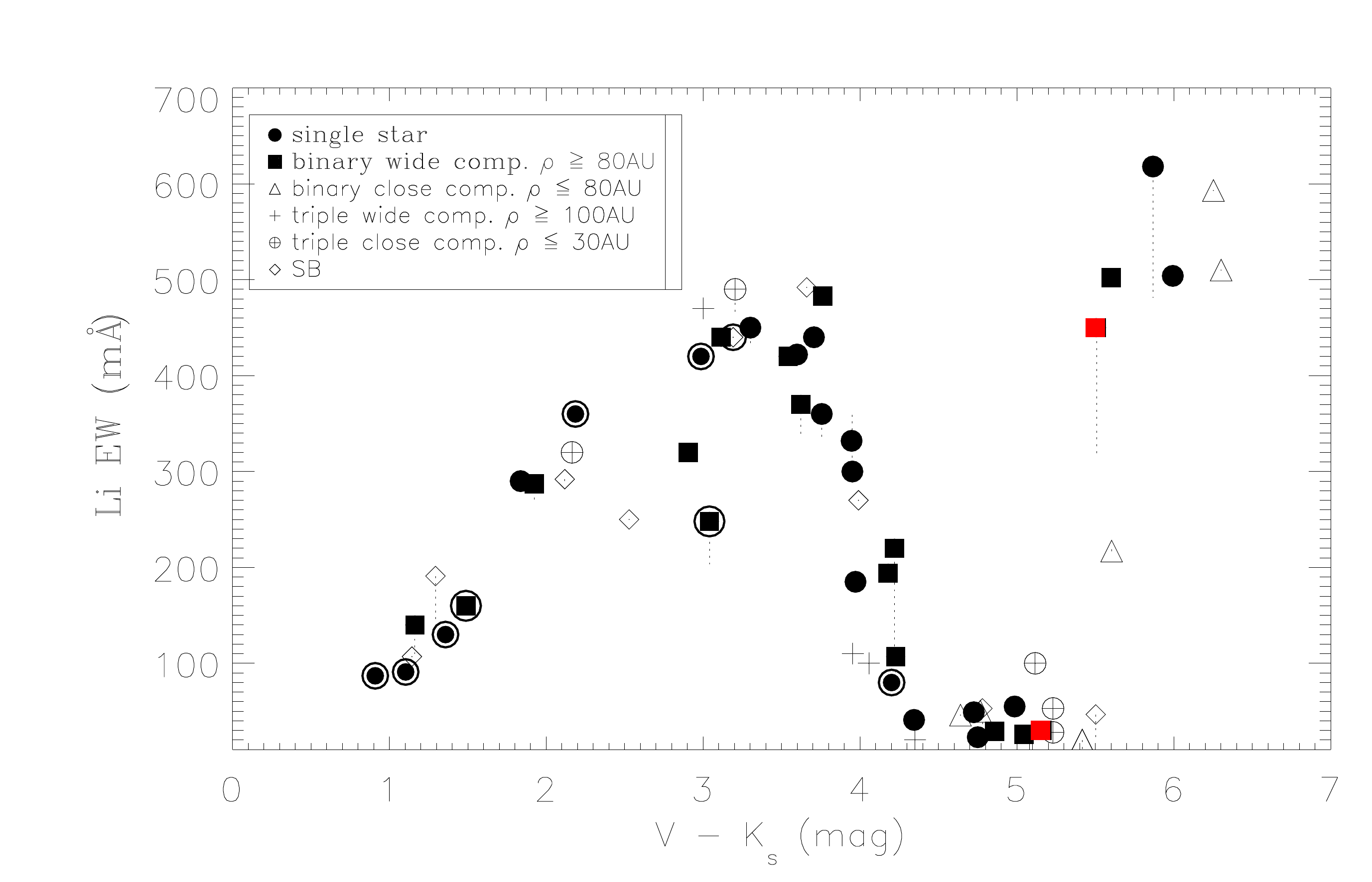} 
\includegraphics[width=50mm,height=80mm,angle=90,trim= 0 0 0 0]{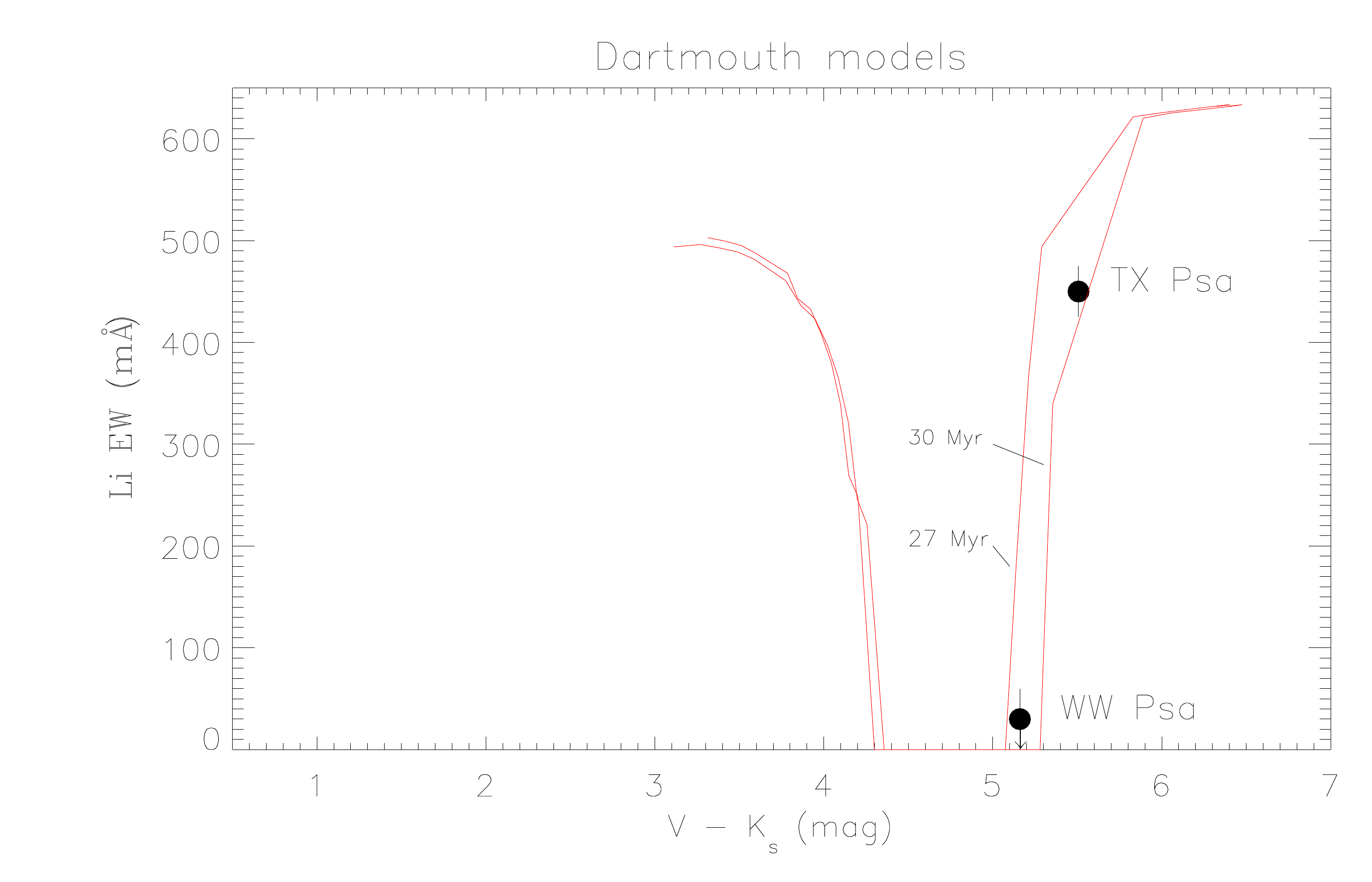} 
\end{minipage}
\caption{\label{li} \it Left panel: \rm Distribution of Li EW among members of the $\beta$ Pictoris association with WW Psa and TX Psa marked in red. \it Right panel: \rm Li abundances for WW Psa and TX Psa with   Li depletion models (solid lines) taken from Feiden (2016) corresponding to the ages of 27\,Myr and 30\,Myr. }
\vspace{0cm}
\end{figure*}

The Li abundance is the first age dependent quantity that we exploit to infer the age of  WW Psa and TX Psa.  
To date, the fit of the LDB represents the most reliable absolute age-dating technique.
\citet{Messina16a} measured an age of 25$\pm$3\,Myr for the $\beta$ Pictoris association by fitting the hot side of the
LDB, using the Dartmouth models of \citet{Feiden16} that include the effects of magnetic fields. 
Whereas in \citet{Messina16a} all bona-fide members were considered, WW Psa and TX Psa included, in 
the present investigation we check whether the ages of the individual stars WW Psa 
and TX Psa are in agreement or not with the age inferred for the whole association. \\
In the left panel of Fig.\,\ref{li}, we note
that WW Psa is in the region of the Li EW distribution occupied by completely depleted stars, whereas the lower-mass 
component TX Psa is in the region occupied by un-depleted stars. In the right panel of Fig.\,\ref{li}, we compare the Li EW of 
both components with the predictions of the Dartmouth evolutionary models. These models were computed for solar metallicity and for an equipartition magnetic field strength in the range 2500 $<$ $\langle$B$f_{eq}$$\rangle$ $<$ 3000\,G. Model effective temperatures were transformed into V$-$K$_s$ colors using the empirical T$_{\rm eff}$-V$-$K$_s$ relation from \citet{Pecaut13} valid for young 5--30\,Myr stars.  Then, to make the comparison with the observations, we transformed the model Li abundance into Li EW. For this purpose, we have used the curves of growth from \citet{Zapatero02}. They are valid in the effective temperature range 2600 $<$ T$_{\rm eff}$ $<$ 4100\,K and for 1.0 $<$ A(Li) $<$ 3.4\footnote{A more detailed discussion on the modeling with the \citet{Feiden16} and the \citet{Baraffe15} models is given in \citet{Messina16a}.}.
We find that the observed Li EWs are best fitted by evolutionary models in the age range from 27\,Myr to 30\,Myr. We refer the reader to \citet{Messina16a} for a more detailed 
discussion about the LDB modeling.\\
This age inferred from LDB fitting is slightly older but still comparable within the uncertainties with the age of the whole association. \\
 However, it is worth noticing the extreme model dependency of such ages. In fact, if one instead adopts non-magnetic models then a much younger age is derived (see e.g. \citealt{Messina16a}). \\ \rm
We note that rotation has a key role in the Li depletion mechanism. First, \citet{Soderblom93}) using the projected rotational velocity of 125-Myr Pleiades members found strong evidence that fast rotators are less depleted than slow rotators. Similar results have been recently found by \citet{Bouvier16} in the young 4-Myr \object{NGC\,2264} open cluster. 
The same Li depletion-rotation connection has been found by \citet{Messina16a}  among the members of the $\beta$ Pictoris association.
However, in that study such correlation is  only \rm well established for the $\beta$ Pictoris members  on \rm the hot side of the LDB. 
On the contrary, in the Li gap and in the cool side of the LDB, where WW Psa and TX Psa are positioned,  the significance of this correlation is not large.
 Therefore, the possible impact of rotation on the Li EW difference found between WW Psa and TX Psa cannot yet be firmly established and the fitted values of Li EW are the observed ones, not decorrelated from rotation. It is worth noting that contrary to the observational evidence,  theoretical studies (\citealt{Eggenberger12}) predict 
that the inclusion of rotation results in a global increase of the lithium depletion during the pre-main-sequence. \rm 
\rm

\subsection{Age from stellar rotation}
 
\begin{figure*}
\begin{minipage}{20cm}
\includegraphics[width=100mm,height=160mm,angle=90,trim= 0 0 0 0]{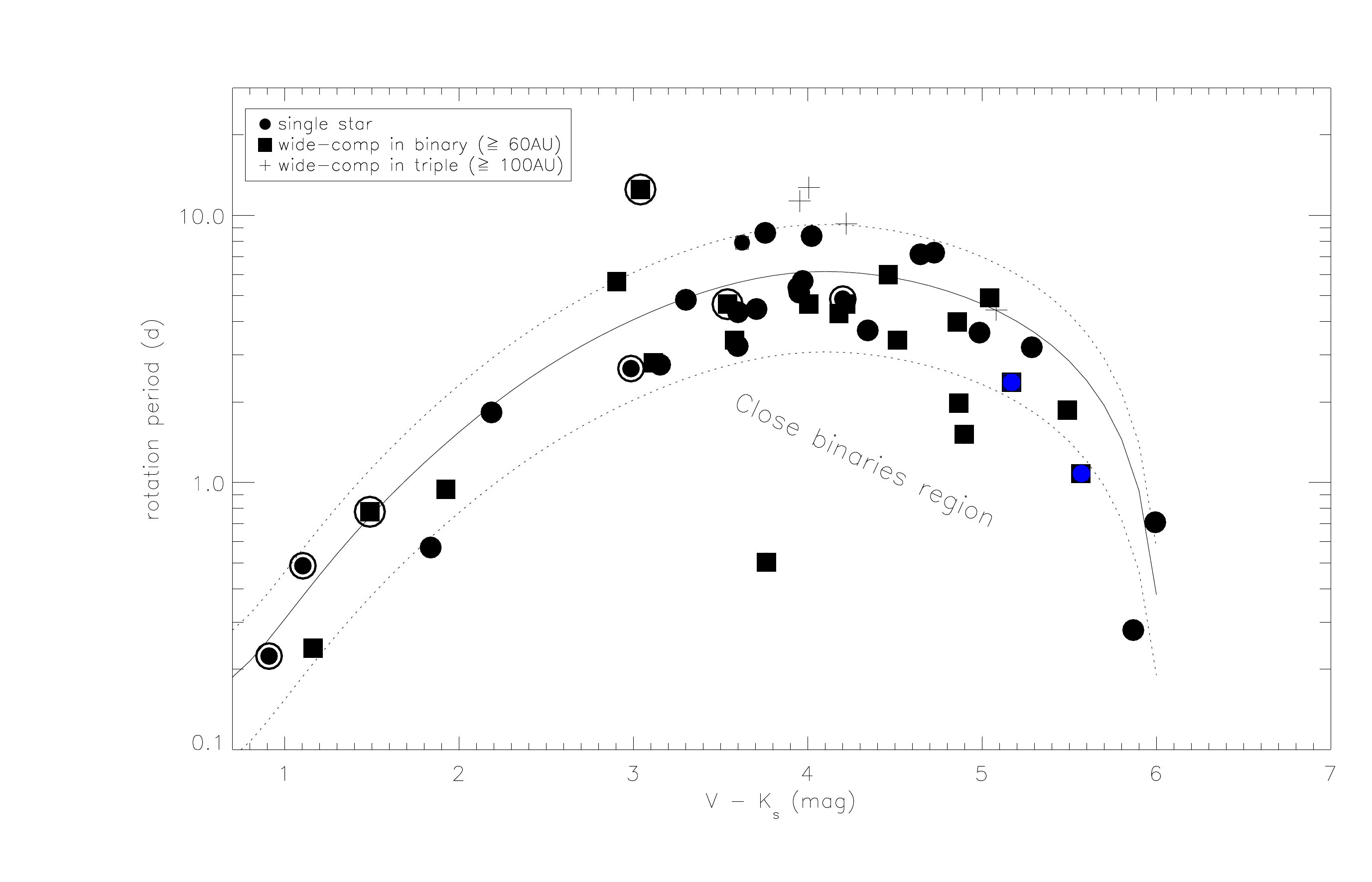} 
\end{minipage}
\caption{\label{distri_period}Distribution of rotation periods of single stars and components of wide binary/triple systems in $\beta$ Pictoris association. Circled symbols are stars hosting a debris disc. The filled blue squares indicate the positions of WW Psa and TX Psa. Solid and dotted lines represent polynomial fits to the median period distribution and to its upper and lower boundaries (from Messina et al. in preparation).}
\vspace{0cm}
\end{figure*}

Stellar rotation is the second age dependent parameter that we use to check if  the rotation periods of WW Psa and TX Psa are consistent with those of other members of the $\beta$ Pictoris association. \rm
Low-mass stars in young clusters/associations exhibit a distribution
of rotation periods that primarily depends on age, mass, and among other factors on the initial rotation period and the  disc lifetime. 
Starting from the higher-mass F and G stars, the width of the period distribution progressively decreases as far as the age increases (see, e.g., \citealt{Mamajek08}), by an age of about 0.6\,Gyr a one-to-one correspondence is reached between mass and rotation period (see, e.g., \citealt{Delorme11}).\\
This means that at the age of $\beta$ Pictoris the distribution of rotation periods overlaps with the period distributions of other clusters/associations either younger or older by  few tens of Myr. 
In contrast to the technique based on the LDB, owing to the width of period distribution at any given mass and  the uncertainty associated with the age of the benchmark association/clusters, a precise absolute age measurement of WW Psa and TX Psa based on  rotation period is not possible. Nonetheless, we can check whether their rotation periods fall within
the distribution of all other bona-fide members or are outliers.\\
In Fig.\,\ref{distri_period}, we plot the period distribution of single stars and wide (separation $>$ 80\,AU\footnote{Messina et al. (in preparation) found that components of binary and multiple systems with separation larger than 80\,AU have a rotation period distribution indistinguishable from that of single stars.}) components of binary/triple systems that are all members of the $\beta$ Pictoris association. Rotation periods and single/binary nature are retrieved from the catalogue of rotation periods of $\beta$ Pic members \citep{Messina16b}. 
We see that the rotation periods of WW Psa and TX Psa fall on the distribution's lower boundary separating slow rotating single stars and wide components of binary/multiple
systems from fast rotating close binary systems (separation $<$ 80\,AU). Considering that the radii of WW Psa and TX Psa are still contracting and, consequently, their rotation periods are spinning up, the rotation tells us that the system may have an age slightly older but still compatible with that of the $\beta$ Pictoris association, if they are effectively single stars. However, 
since WW Psa and TX Psa also fall on the upper boundary of the distribution of close binary/multiple members,  we may be dealing with two components that are themselves unresolved close binaries.\\
Therefore, the rotation periods of WW Psa and TX Psa are compatible with those of other bona fide members, although there is some hint of a possible unresolved binary nature
of both components. 
\rm

 \subsection{Age from isochrone fitting}
\begin{figure*}
\begin{minipage}{20cm}
\includegraphics[width=50mm,height=80mm,angle=90,trim= 0 0 0 0]{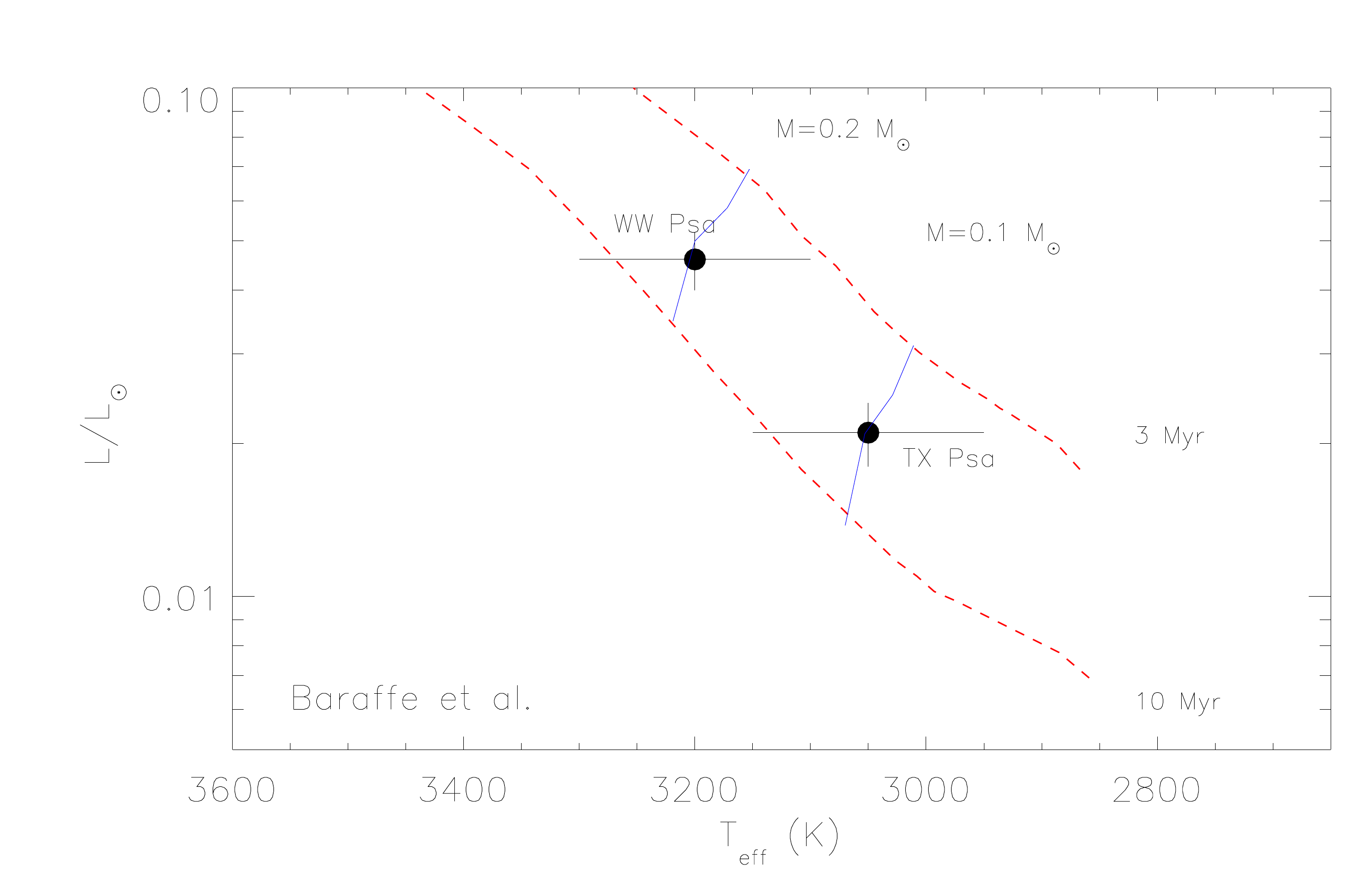} 
\includegraphics[width=50mm,height=80mm,angle=90,trim= 0 0 0 0]{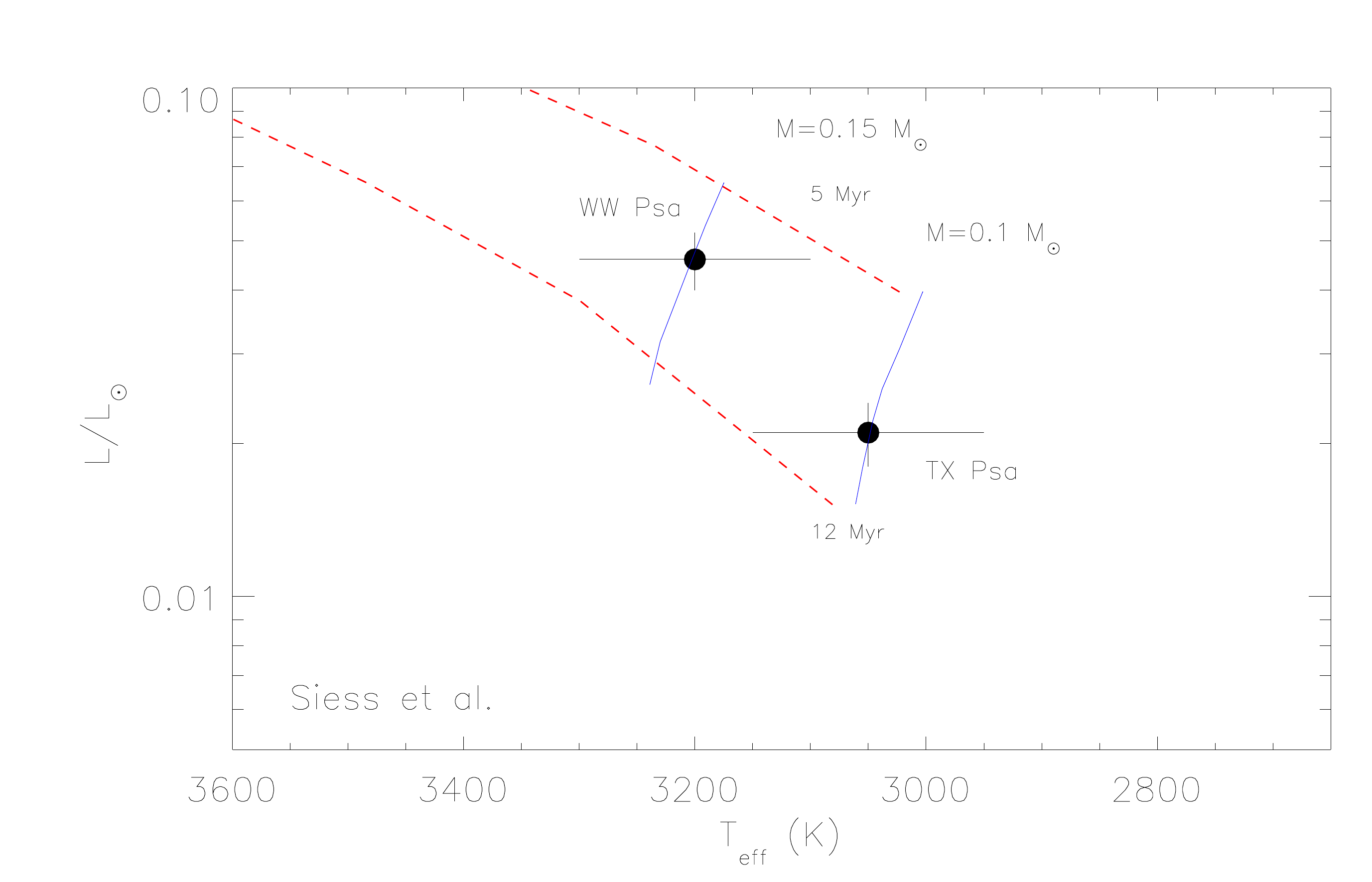} \\
\includegraphics[width=50mm,height=80mm,angle=90,trim= 0 0 0 0]{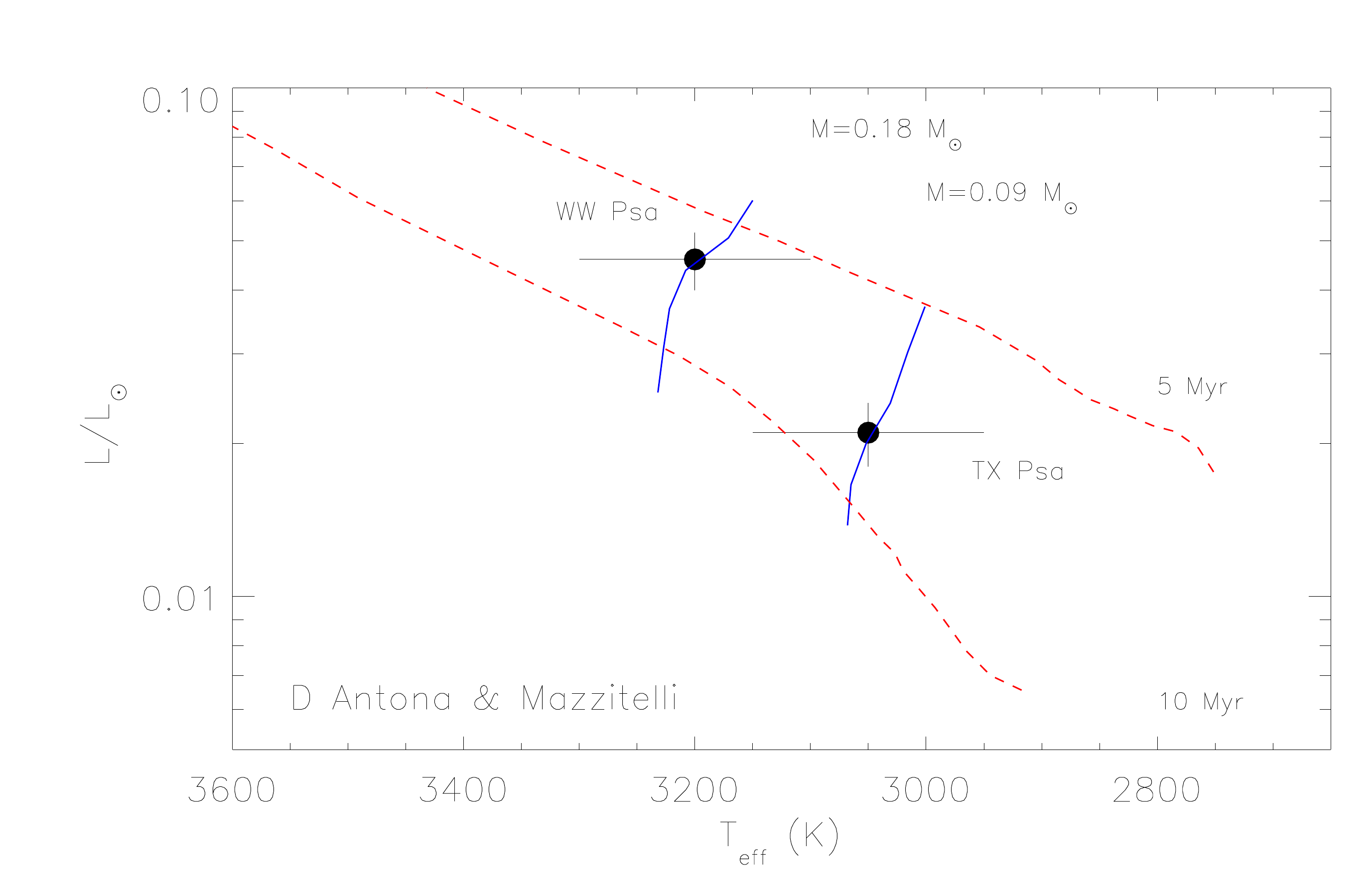} 
\includegraphics[width=50mm,height=80mm,angle=90,trim= 0 0 0 0]{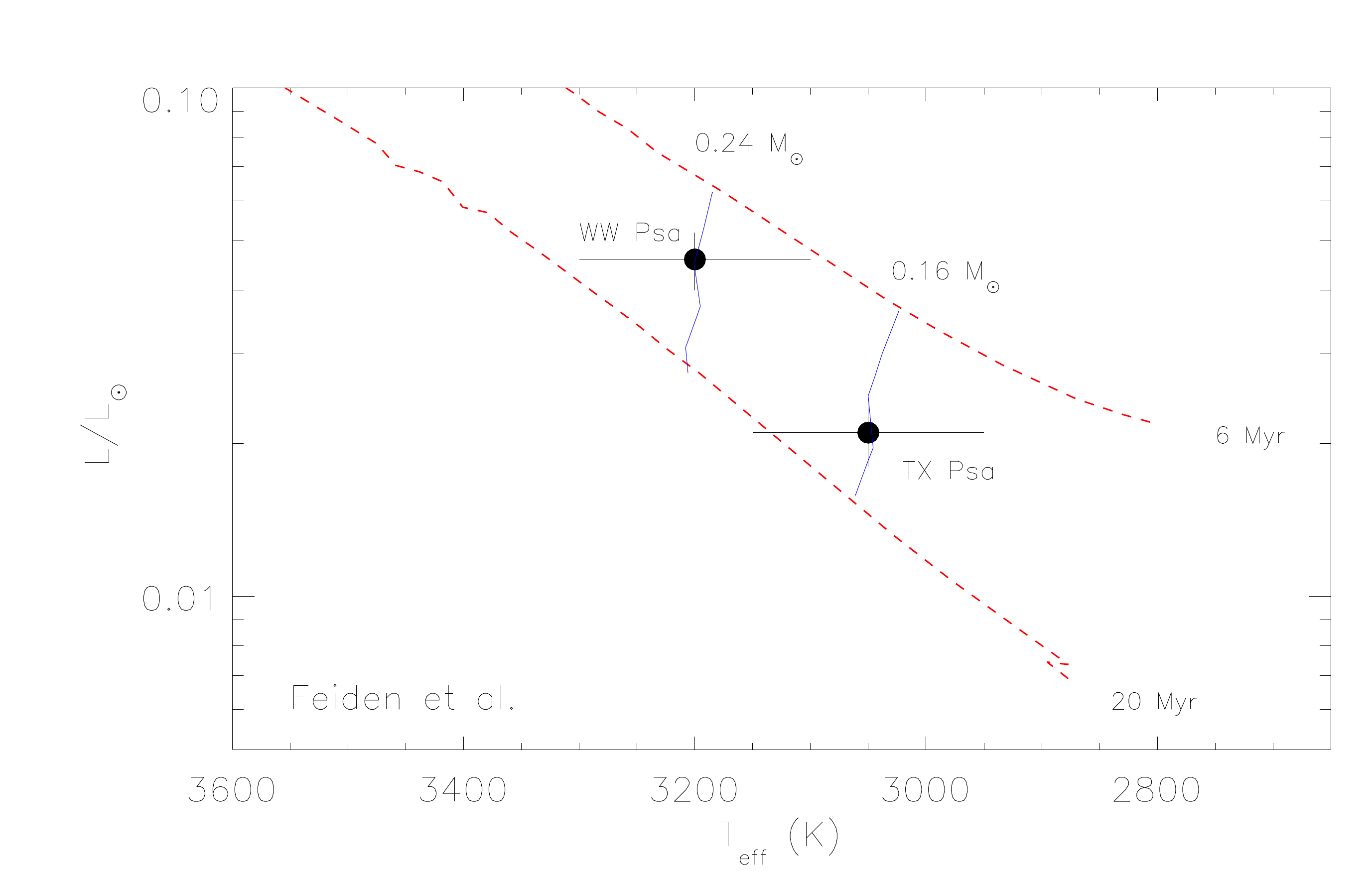} 
\end{minipage}
\caption{\label{hr}HR diagrams. Dashed lines are isochrones whereas  blue solid lines are the  evolutionary tracks. Models are from Baraffe et al. (2015) (top left panel), Siess et al. (2000) (top right panel), D'Antona \& Mazzitelli (1997) (bottom left panel), and Feiden et al. (2016) (bottom right panel).}
\vspace{0cm}
\end{figure*}

 Luminosity and effective temperature are the last age dependent quantities that we fit with isochrones to infer the age of WW Psa and TX Psa. The observed magnitude, distance, and bolometric correction are used to derive the luminosity    that, together with the effective temperature, \rm allows us to compare the positions of both components on the   Hertzsprung-Russell (HR) \rm diagram with  isochrones  from different evolutionary models. \rm We   adopt \rm for WW Psa and TX Psa the most recent trigonometric distance d = 20.75$\pm$0.25\,pc provided by GAIA (Gaia collaboration \citeyear{Gaia16}). Using their brightest (and presumably unspotted) observed magnitudes V = 12.10$\pm$0.02\,mag and V = 13.35$\pm$0.02\,mag (derived from ASAS time series; \citealt{Pojmanski02}), and the bolometric corrections BC$_{\rm V}$ = $-$2.43$\pm$0.05\,mag  and BC$_{\rm V}$ = $-$2.82$\pm$0.05\,mag, taken from \citet{Pecaut13} and   corresponding to  effective temperatures inferred from the SED fits (see Sect.\,4), \rm
we infer the luminosities  L = 0.046$\pm$0.006\,L$_\odot$   and L = 0.021$\pm$0.003\,L$_\odot$    respectively (adopting M$_{\rm bol,_\odot}$ = 4.74\,mag \rm and T$_{\rm eff_\odot}$ = 5777\,K from \citealt{Cox00}). \\
 \rm In Fig.\,\ref{hr},  WW Psa and TX Psa  are compared with a grid of isochrones and mass evolutionary tracks taken from
 different evolutionary models. We find that  both components have an age in the range 3--10\,Myr according to  \citet{Baraffe15}, in the range 5--12\,Myr according to \citet{Siess00}, in the range 5--10\,Myr accorging to \citet{D'Antona97}, and in the range 6--20\,Myr according to  \citet{Feiden16}.  \rm 
Only the Dartmouth models of  \citet{Feiden16}, where also the effects of magnetic fields are included, predict an older age but not older than 20\,Myr. The inferred masses, as typed in  Fig.\,\ref{hr}, are in the range 0.15--0.24\,M$_\odot$ for WW Psa and in the range 0.09--0.16\,M$_\odot$ for TX Psa, depending on the adopted model. \\
All models provide ages significantly younger than the quoted 25$\pm$3\,Myr age of the $\beta$ Pic association. The discrepancy is only partly mitigated in the Dartmouth models, showing that
the magnetic fields play some role in these very active stars through inhibition of stellar convection, which results in stars with larger radius and higher luminosity at a given age, with respect to non-active stars. 
This residual age discrepancy suggests that some other physics, such as \rm starspots (e.g., \citealt{Somers15}; \citealt{Jackson14})  is probably missing in the stellar evolution models.   It is worth noting that the position in the HR diagram \rm of WW Psa and TX Psa, which makes them to appear over luminous with respect to the  adopted age of 25$\pm$3 Myr, \rm is not peculiar of these stars but also shared by other $\beta$ Pictoris members of similar effective temperatures (see, e.g., Fig.\,4 in \citealt{Messina16a}).\\ 
\rm

\subsubsection{Unresolved binary hypothesis}
We investigate the possibility that WW Psa and TX Psa may be unresolved binary stars using the available information from RV measurements and imaging.
The radial velocities of both components have been monitored quite extensively.
\citet{Elliott14} measured the RV for both components. Here, we report the updated values $<$RV$>$ = 0.95$\pm$1.35\,km\,s$^{-1}$ and $<$RV$>$ = 3.2$\pm$1.0\,km\,s$^{-1}$ for WW Psa and TX Psa, respectively, which were re-calculated using an M dwarf mask that improved the  cross-correlation-function \rm  fit with respect to \citet{Elliott14}.   Briefly, 
the CORAVEL-type numerical mask was originally created from a stellar spectrum. The spectrum is converted to a numerical mask so that the continuum has a value of 0 and the absorption lines have the value 1 at their peak.  The mask is then convolved with the observed spectrum of desired target to create the cross correlation function.
The quoted uncertainties  are calculated from the standard deviation of individual radial velocity measurements, as opposed to direct measurement uncertainties.  It is true that the associated uncertainties are larger than those presented in \citet{Elliott14}.  However, from the recalculation of the radial velocities using the M-type template there was more variability in the resultant values.  This could be explained by a better match of the numerical mask with the observed spectra combined with the effect of star spots that can mimic radial velocity variance (see \citealt{Lagrange13}). \\ \rm  
 From the literature we retrieved the following RV measurements for WW Psa:   RV = 2.2\,km\,s$^{-1}$ (\citealt{Torres06});  the average values  $<$RV$>$ = 3.087$\pm$0.134\,km\,s$^{-1}$ (\citealt{Bailey12}); RV = 3.2$\pm$0.5\,km\,s$^{-1}$ (\citealt{Shkolnik12}) and RV = 2.9$\pm$0.6\,km\,s$^{-1}$ (\citealt{Malo14a}). Whereas for TX Psa we retrieved: RV = 2.4\,km\,s$^{-1}$ (\citealt{Torres06}); $<$RV$>$ = 2.031$\pm$0.168\,km\,s$^{-1}$ (\citealt{Bailey12});  RV = 2.2\,km\,s$^{-1}$  (\citealt{Delorme12}). \rm   Based on these measurements, \rm we have the following average values $\langle$RV$\rangle$ = 2.42$\pm$0.84\,km\,s$^{-1}$ for WW Psa and  $\langle$RV$\rangle$ = 2.54$\pm$0.60\,km\,s$^{-1}$  \rm and 
 for TX Psa. \rm Both standard deviations are larger than the uncertainties associated with the single measurements and, as discussed earlier, also larger  than the expected RV variations due to jitters generated by the magnetic activity, which in the infrared are of the order of 0.15\,km\,s$^{-1}$. However, the possibility of Keplerian origin of the  low-amplitude RV variations remains quite marginal.   In fact, \citet{Elliott14}, from a study of hundreds of stars, find a standard deviation of 3\,km\,s$^{-1}$  as the criterion for significant radial velocity variation.\\ \rm 
Both components have also been observed with high-spatial resolution high-contrast imaging. \citet{Delorme12} observed both components with NACO/VLT finding no presence of companions at mass ratios larger than about 0.1 and at a projected separation larger than about 10\,AU (or less for TX Psa). Similar results were achieved by \citet{Biller13}  and \citet{Elliott15}    who observed TX Psa with NICI/Gemini and NACO/VLT, respectively. \rm

\begin{figure*}
\begin{minipage}{20cm}
\includegraphics[width=80mm,height=60mm,angle=0,trim= 0 0 0 0]{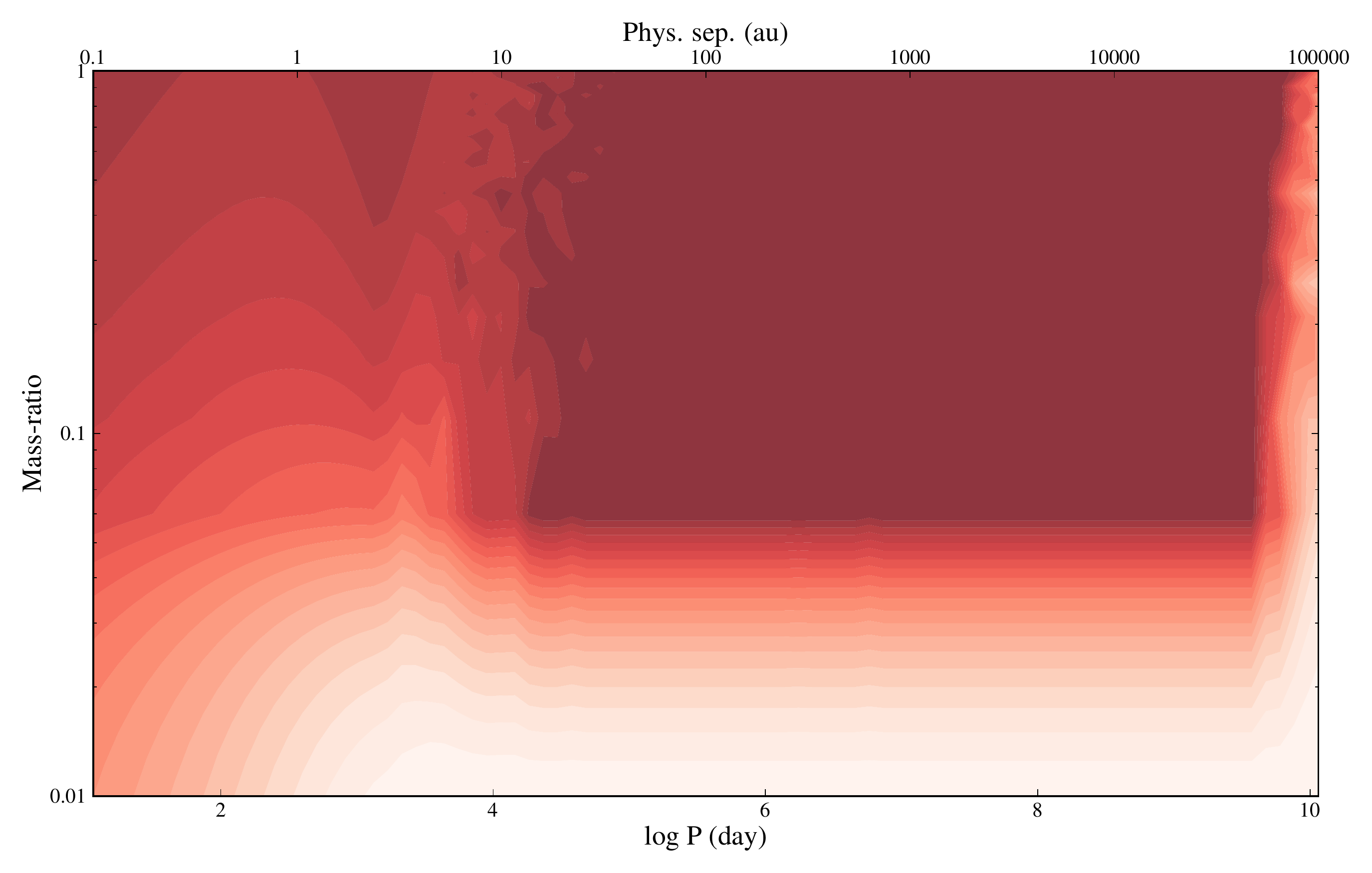} 
\includegraphics[width=80mm,height=60mm,angle=0,trim= 0 0 0 0]{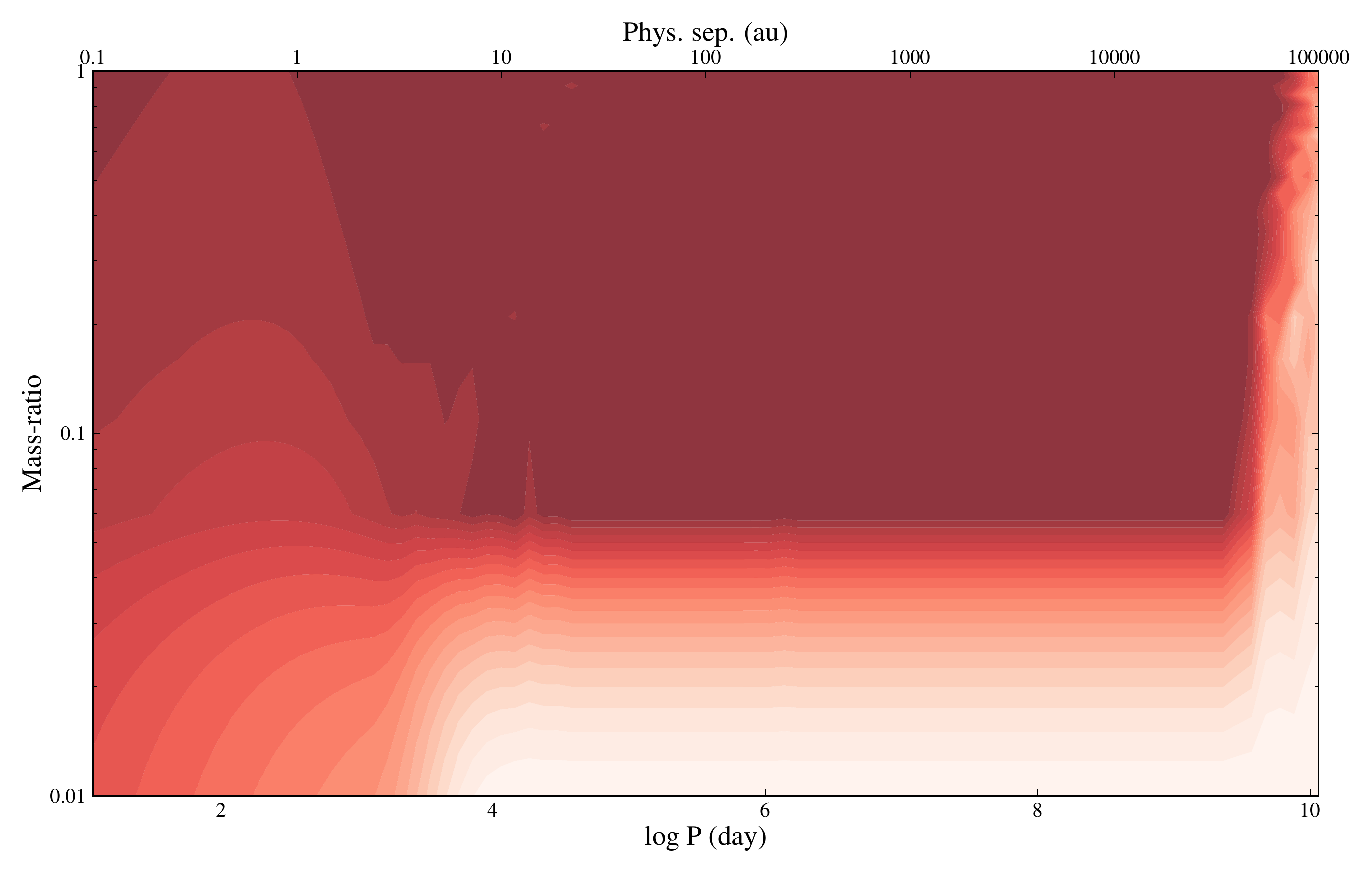} 
\end{minipage}
\caption{\label{simul}  Companion detection probabilities (contours from red (100\%), to white (0\%), in 5\% colour levels) for WW Psa (left) and TX Psa (right) with mass ratio versus orbital period (day) / physical separation (AU). Note that both x- and y-scales are logarithmic. \rm}
\vspace{0cm}
\end{figure*}

In Fig.\,\ref{simul}, we plot  the detection limits from simulated binaries (dark red is 100\% detection, white is 0\%) as a function of separation and mass ratio for WW Psa (left) and TX Psa (right). The companions would therefore have to have a mass-ratio $<$ 0.1 (or a separation smaller than $\sim$10\,AU for WW Psa and smaller than $\sim$1\,AU for TX Psa) for them to be missed so far, whereas it is unlikely that the companions are equal mass.  
Detection limits from RV measurements were derived using multi-epoch observations following \citet{Tokovinin14}. For high-contrast imaging, we converted the angular separation versus contrast, to physical separation versus mass-ratio, using the targetÕs distance and the evolutionary models of \citet{Baraffe15}. We used the detection limits described in \citet{Elliott16}; combining 2MASS photometry (\citealt{Cutri03}) and proper motions (UCAC4, PPMXL, NOMAD: \citealt{Zacharias13}; \citealt{Roeser10}; \citealt{Zacharias04}), for the widest parameter space ($>$ 3$^{\prime\prime}$).  \rm

\begin{figure*}
\begin{minipage}{20cm}
\includegraphics[width=50mm,height=80mm,angle=90,trim= 0 0 0 0]{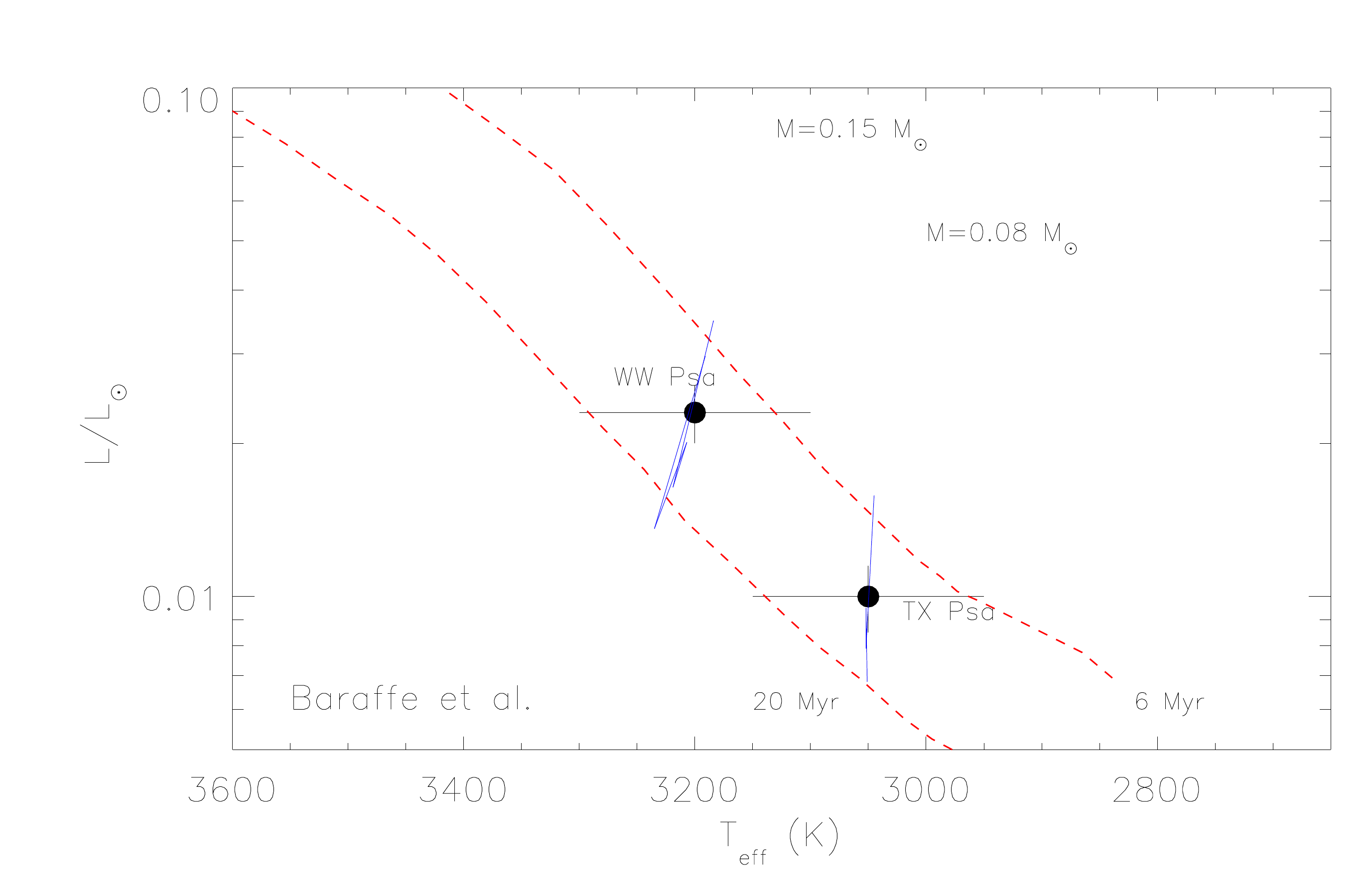} 
\includegraphics[width=50mm,height=80mm,angle=90,trim= 0 0 0 0]{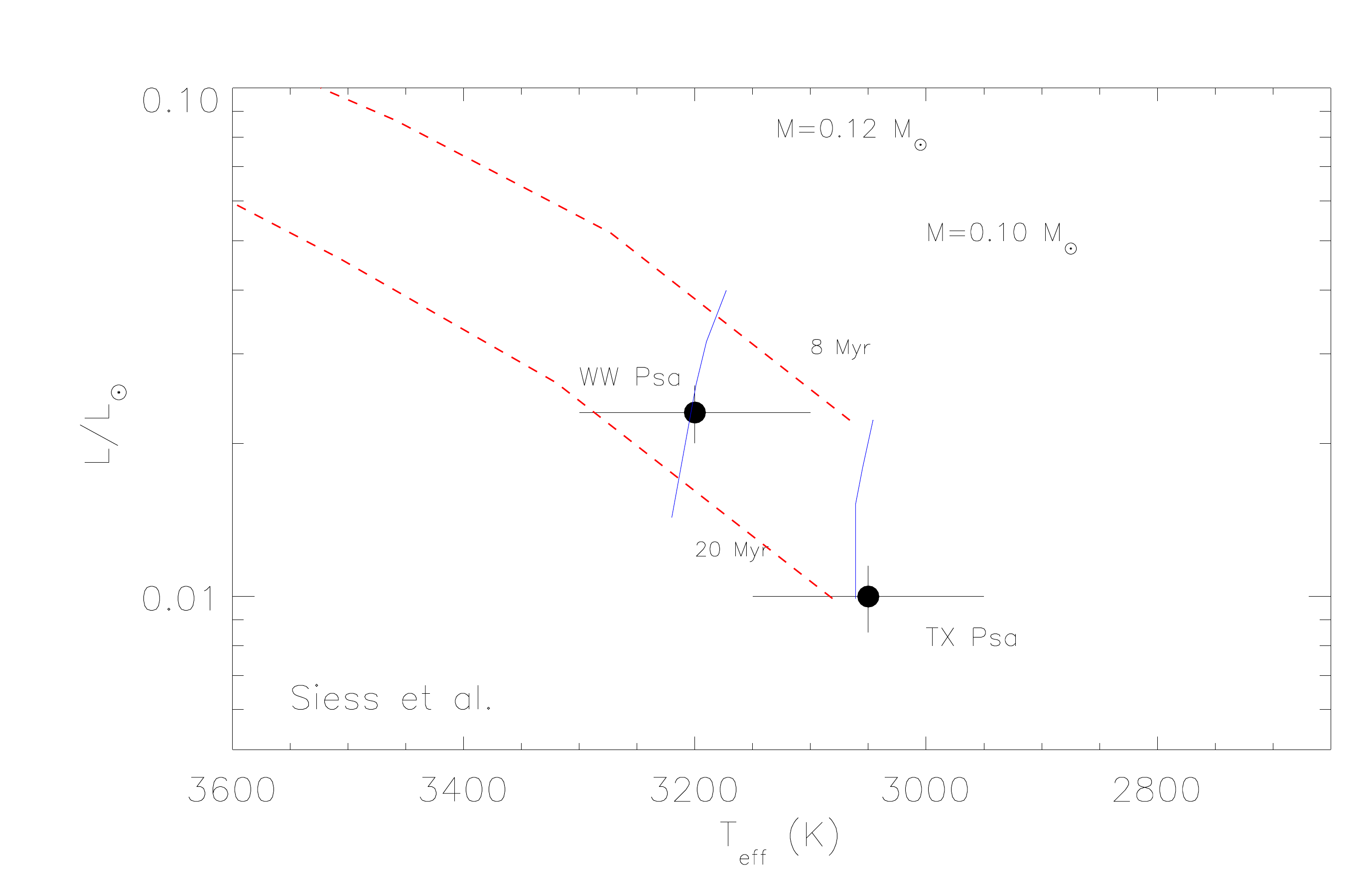} \\
\includegraphics[width=50mm,height=80mm,angle=90,trim= 0 0 0 0]{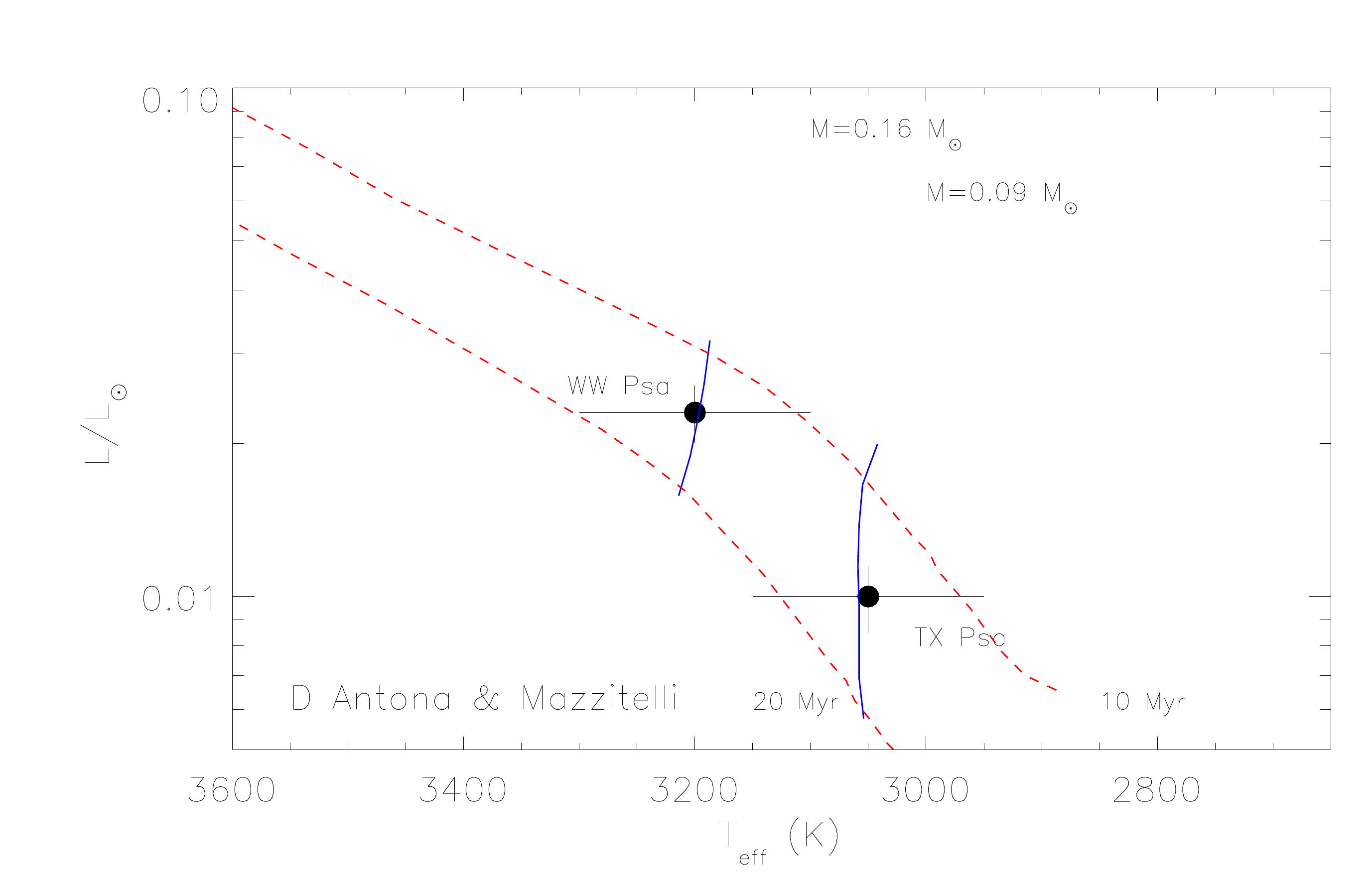} 
\includegraphics[width=50mm,height=80mm,angle=90,trim= 0 0 0 0]{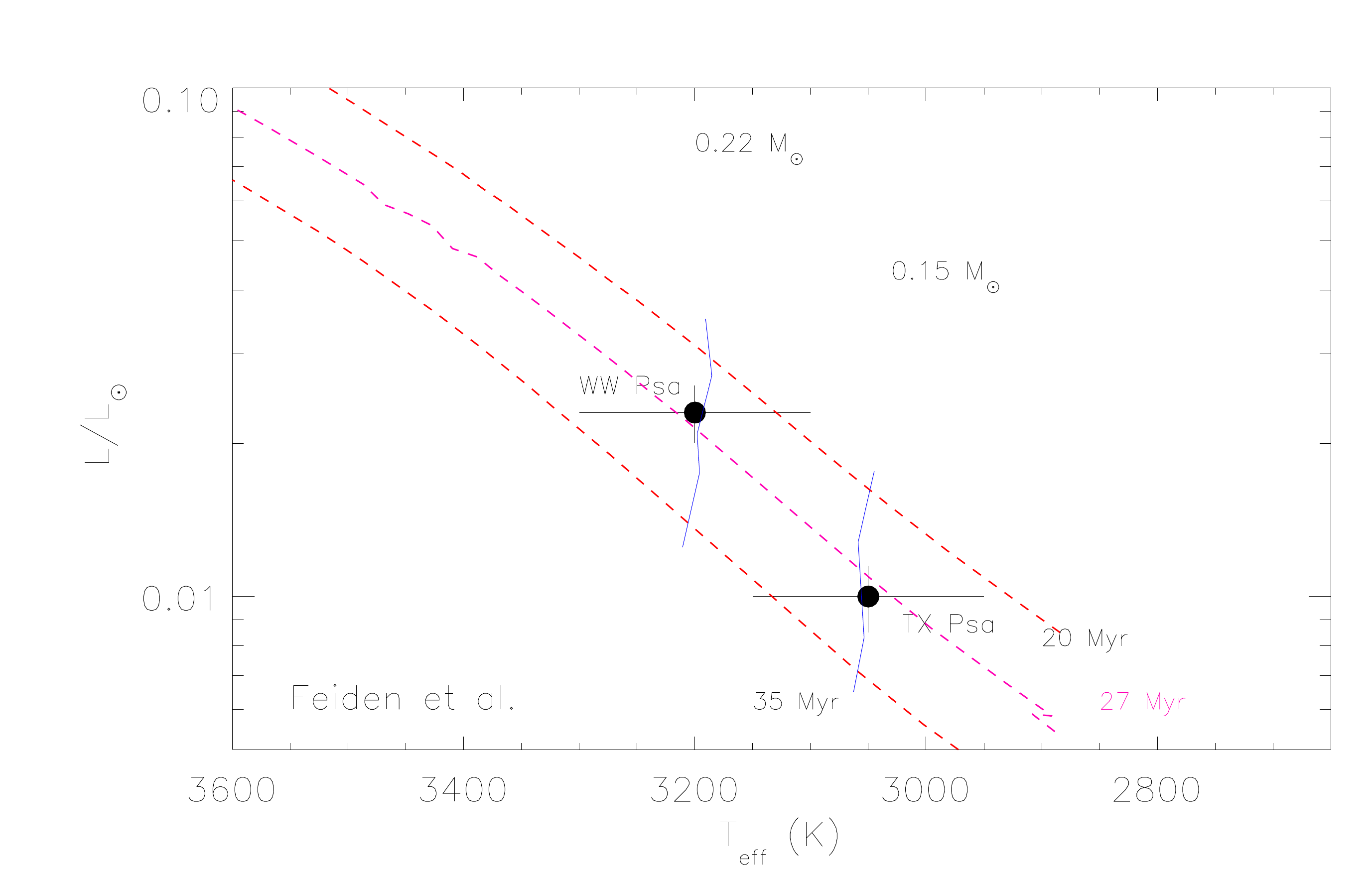} 
\end{minipage}
\caption{\label{hr_bis}The same as in Fig.\,\ref{hr}, but where WW Psa and TX Psa are considered as primary components
of unresolved systems with equal-mass components.}
\vspace{0cm}
\end{figure*}

The single nature of WW Psa and TX Psa is supported also by
 considerations concerning the position in the HR diagram and the photometric behaviour.\\ 
\rm
 In Fig.\,\ref{hr_bis} we plot the luminosities of WW Psa and TX Psa, after correction of the observed magnitude by $\Delta$V =   0.75\,mag  assuming both stars consist of unresolved equal-mass components. \rm  The isochronal fitting was carried out using the same models mentioned in the previous Sect.\,5.3. We find that depending on which  model is adopted among \citet{Baraffe15}, \citet{Siess00}, and \citet{D'Antona97}, \rm the ages of both components
can be older than about 6\,Myr but not older than 20\,Myr. The Dartmouth models provide the best fit with an age of 27$\pm$7\,Myr, in agreement with the estimated age of the association, despite the quite large uncertainty.  \\
 Therefore, the ages of these stars estimated from isochronal fitting with Baraffe et al., Siess et al., and D'Antona \& Mazzitelli models are still much younger than the age of the $\beta$ Pictoris association. An agreement is found only with the magnetic Dartmouth models, but with the unlikely assumption of equal-mass components. Assuming a mass ratio $<$ 0.1, also  the Dartmouth models predict too young ages. 
\rm
Concerning the photometric behaviour, \rm equal-mass components would exhibit similar  levels of magnetic activity and likely would equally contribute to the observed variability making both rotational periods to appear in the periodogram. On the contrary, we detected only one period. Therefore, the secondary undetected components should have lower luminosities giving a negligible contribution to the observed variability. We note that, if the secondary components have lower masses and, therefore, redder colors, then the primaries should be slightly
bluer than observed. This circumstance would shift the position of WW Psa and TX Psa in the period-color diagram  towards bluer V$-$K$_s$ color, positioning both stars in the region occupied by close binary systems (see Fig.\,7). That would better fit the rotational properties of both components that are on the boundary between single and close binaries in the period-color diagram. \\
To summarize, assuming that WW Psa and TX Psa are unresolved close binaries with mass-ratio $<$ 0.1 results in better agreement with the quoted age of the $\beta$ Pictoris association. However, even in this hypothesis both stars remain more luminous/cooler than predicted by a 25-Myr isochrone.

\subsection{Effects by spots hypothesis}
Both stars are very active as   demonstrated \rm by the significant NUV and FUV flux excesses and by the photometric variability. 
  The maximum amplitude of the photometric variability measured in our campaign is  \rm $\Delta$V = 0.09\,mag and $\Delta$V = 0.07\,mag for WW Psa and TX Psa, respectively. \\
These amplitudes provide a lower limit to the fraction of stellar photosphere covered by spots,    since additional spots evenly distributed in longitude might be present, without  contributing to the light rotational modulation but increasing the total covering fraction. \rm  The inclination of the stellar rotation axis can also keep the observed variability at a low level, even in the cases when the covering fraction is significant. \\
For WW Psa we derive (from luminosity and effective temperature) a stellar radius   R = 0.70$\pm$0.13\,R$_\odot$ \rm that combined with the stellar rotation period P = 2.37$\pm$0.01\,d provides the inclination of the stellar rotation axis    $i^{\circ}$ =  53$^{+22}_{-12}$ \rm (adopting $<v\sin{i}>$ = 12$\pm$1.1 \,km\,s$^{-1}$; see Table 1)\rm. \\
Similarly, for TX Psa we derive a stellar radius    R = 0.52$\pm$0.10 R$_\odot$ \rm  that combined with the rotation period P = 1.086$\pm$0.003\,d gives an inclination of the stellar rotation axis    $i^\circ$ = 60$^{+20}_{-18}$ \rm  (adopting $<v\sin{i}>$ = 21$\pm$3 \,km\,s$^{-1}$;  see Table 1\rm). Uncertainties on  radius and inclination are computed according to the propagation of error on all involved quantities.\\ 
The two components of this physical pair are found to have similar inclinations of their rotation axes that   act \rm to reduce any spot's visibility and, consequently, the amplitude of the rotational modulation. Therefore,  
both components may have a significant fraction of their photospheres  covered by spots while still producing a light curve amplitude not larger than $\Delta$V = 0.09\,mag and $\Delta$V = 0.07\,mag for each component, respectively.\\ \rm
In a recent series of papers by Somers \& Pinsonneault (see, e.g., \citealt{Somers15})  the impact of starspot activity on the mass and age of   pre-main-sequence \rm stars as inferred from isochronal fitting were investigated. Their models demonstrate that   when a significant fraction of the surface is covered in starspots, \rm   stars are displaced  towards lower masses and effective temperatures in a HR diagram. As a consequence, spotted stars appear younger and less massive than they effectively are. According to their predictions, in the case of WW Psa, the derived mass should be increased by  a factor from 1.3 to 2.0 and its age by a factor from 1.8 to 6.0 if the covering fraction ranges from 15\% to 50\% at an age of 10 Myr. Such correction factors decrease at increasing age.\\
  The use of the Dartmouth models that incorporate the effects of magnetic fields mitigates the age discrepancy, indicating that magnetic activity in these stars might plays some role \rm
in altering basic parameters such as effective temperature and luminosity. However, these models do not provide the expected ages of these very red stars.\\

\section{Conclusions}
We have carried out a multi-site multi-epoch   photometric monitoring campaign \rm of WW Psa and TX Psa. This monitoring has allowed us to confirm the already known rotation period of WW Psa and to measure for the first time the rotation period of TX Psa (P = 1.086\,d).   Moreover, we estimated  effective temperatures of both components  by fitting stellar atmospheric models to the SEDs using archival  optical and near-IR photometry. \rm 
 We explore if the measured rotation periods, the Li EW available from the literature, and position on the HR diagram, which are all age dependent, are consistent with the age of 25\,Myr of the $\beta$ Pic association inferred by, e.g., \citet{Messina16a} with the LDB fitting technique.\rm 
We find that both the rotation periods, which are compared with the period distribution of the other association bona fide members, and the Li EWs, which are compared with the evolutionary models of \citet{Feiden16}, are similar to the values measured in equal-mass members of the $\beta$ Pictoris association. \rm On the contrary, the isochronal fitting of luminosities and effective temperatures in the HR diagram infer an age for this system significantly younger than the quoted age of 25\,Myr. We probe the possibility that the discrepancy may arise from unresolved binary nature of both components. However, this possibility is supported neither by RV studies nor from imaging and simulations nor from the photometric variability behavior.  The use of models that incorporate the effects of magnetic fields significantly mitigates the age discrepancy. However, since the discrepancy
still remains, we conclude that some other physics not taken into account plays a key role in determining the observed luminosity
and temperature effective of these very low mass stars.\\ \rm

{\it Acknowledgements}. Research on stellar activity at INAF- Catania Astrophysical Observatory is supported by MIUR  (Ministero dell'Istruzione, dell'Universit\'a e della Ricerca).  This research has made use of the Simbad database, operated at CDS (Strasbourg, France). This publication makes use of VOSA, developed under the Spanish Virtual Observatory project supported from the Spanish MICINN through grant AyA2011-24052. SM thanks Lison Malo for providing updated values of space and velocity components. We are especially grateful to the Referee whose comments helped us to significantly improve our analysis and the paper quality.

\bibliographystyle{aa.bst} 
\bibliography{mybib} 

\end{document}